\documentclass[a4paper,11pt]{article}
\pdfoutput=1 

\usepackage{jinstpub} 
\usepackage{ulem}
\usepackage{siunitx}
\usepackage{wrapfig}

\usepackage{booktabs}
\usepackage{multirow, makecell}
\usepackage{listings}
\usepackage{hyperref}
\hypersetup{colorlinks=true,linkcolor=blue}

\usepackage{floatrow}
\usepackage{ragged2e}

\usepackage{subfig}
\usepackage{graphicx}

\DeclareSIUnit \s {\second}
\DeclareSIUnit \ns {\nano\second}
\DeclareSIUnit \mus {\micro\second}
\DeclareSIUnit \ms {\milli\second}
\DeclareSIUnit \MB {\mega\byte}
\DeclareSIUnit \GB {\giga\byte}
\DeclareSIUnit \TB {\tera\byte}
\DeclareSIUnit \PB {\peta\byte}
\DeclareSIUnit \Mbps {\mega\bit/\s}
\DeclareSIUnit \Gbps {\giga\bit/\s}
\DeclareSIUnit \Tbps {\tera\bit/\s}
\DeclareSIUnit \Pbps {\peta\bit/\s}
\DeclareSIUnit \kton {\kilo\tonne} 
\DeclareSIUnit \kt {\kilo\tonne}
\DeclareSIUnit \Mt {\mega\tonne}
\DeclareSIUnit \eV {\electronvolt}
\DeclareSIUnit \keV {\kilo\electronvolt}
\DeclareSIUnit \MeV {\mega\electronvolt}
\DeclareSIUnit \GeV {\giga\electronvolt}
\DeclareSIUnit \PeV {\peta\electronvolt}
\DeclareSIUnit \EeV {\exa\electronvolt}
\DeclareSIUnit \ZeV {\zetta\electronvolt}
\DeclareSIUnit \m {\meter}
\DeclareSIUnit \cm {\centi\meter}
\DeclareSIUnit \in {\inchcommand}
\DeclareSIUnit \km {\kilo\meter}
\DeclareSIUnit \kV {\kilo\volt}
\DeclareSIUnit \kW {\kilo\watt}
\DeclareSIUnit \MW {\mega\watt}
\DeclareSIUnit \MHz {\mega\hertz}
\DeclareSIUnit \mrad {\milli\radian}
\DeclareSIUnit \year {years}
\DeclareSIUnit \POT {POT}
\DeclareSIUnit \sig {$\sigma$}
\DeclareSIUnit\parsec{pc}
\DeclareSIUnit\lightyear{ly}
\DeclareSIUnit\foot{ft}
\DeclareSIUnit\ft{ft}
\DeclareSIUnit \ppb{ppb}
\DeclareSIUnit \ppt{ppt}
\DeclareSIUnit \samples{S}
\DeclareSIUnit \pe{PE}
\DeclareSIUnit \mwe{mwe}

\newcommand{\enu}{\E_\enu}

\newcommand{\ts}{\textsuperscript}

\title{\boldmath Convolutional Neural Networks for Shower Energy Prediction in Liquid Argon Time Projection Chambers}

\author[a]{K. Carloni}
\author[a]{N.W. Kamp}
\author[a]{A. Schneider}
\author[a]{J.M. Conrad}

\affiliation[a]{Dept.~of Physics, Massachusetts Institute of Technology, Cambridge, MA 02139, USA}

\emailAdd{kcarloni@mit.edu}
\emailAdd{nwkamp@mit.edu}
\emailAdd{aschn@mit.edu}
\emailAdd{conrad@mit.edu}

\abstract{When electrons with energies of $O(100)~\si\MeV$ pass through a liquid argon time projection chamber (LArTPC), they deposit energy in the form of electromagnetic showers.
Methods to reconstruct the energy of these showers in LArTPCs often rely on the combination of a clustering algorithm and a linear calibration between the shower energy and charge contained in the cluster.
This reconstruction process could be improved through the use of a convolutional neural network (CNN).
Here we discuss the performance of various CNN-based models on simulated LArTPC images, and then compare the best performing models to a typical linear calibration algorithm.
We show that the CNN method is able to address inefficiencies caused by unresponsive wires in LArTPCs and reconstruct a larger fraction of imperfect events to within $\SI{5}\percent$ accuracy compared with the linear algorithm.}

\arxivnumber{2110.10766} 

\begin{document}
\maketitle
\flushbottom

\section{Introduction}
\label{section:intro}

\begin{figure}
    \centering
    \includegraphics[width=\linewidth]{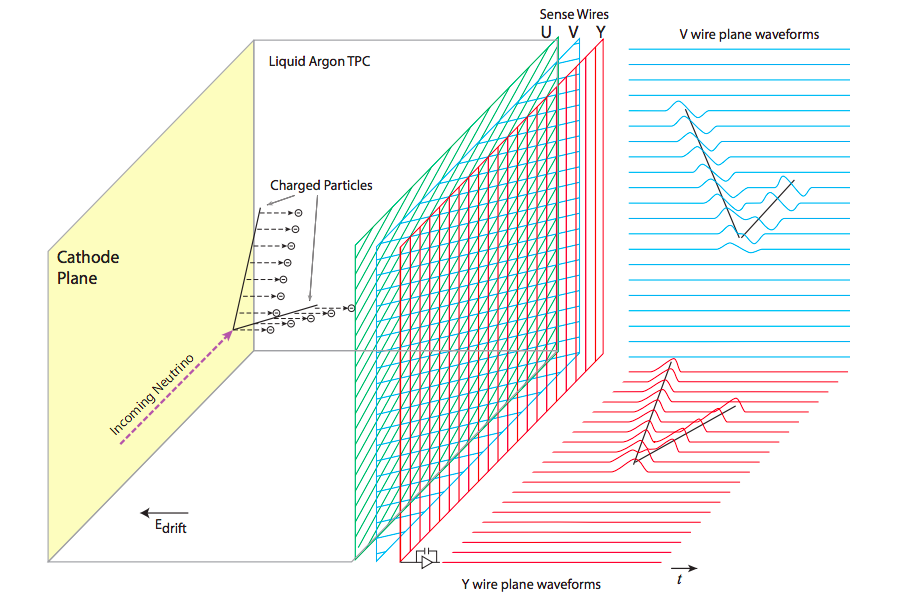}
    \caption{Diagram showing the operational principle behind a LArTPC. Charged particles traverse the argon volume, leaving a trail of ionization electrons which are drifted through an electric field to a series of wire planes. Figure adapted from Ref.~\cite{MicroBooNE:2016pwy}}
    \label{fig:tpc_example}
\end{figure}

Liquid argon time projection chambers (LArTPCs)~\cite{Rubbia:1977zz} consist of a large volume of liquid argon, a drift region within the argon where a static electric field is maintained, and planes of anode wires where charge is measured and collected.
A charged particle traversing the detector will ionize the liquid argon, leaving a trail of ionization electrons along its path.
The static electric field is used to drift these ionization electrons.
The anode planes consist of a series of independent wires on which the arriving charge is measured.
This segmented readout provides two dimensional charge information (per wire and per unit time) from each anode plane.
{Thus, the data coming from a given anode plane can be thought of as an image, in which each ``pixel'' corresponds the amount of charge measured on a single wire over a specified time interval.} 
{Figure~\ref{fig:tpc_example} gives a schematic summarizing the operational principle behind a LArTPC.}\\

{LArTPCs are an increasingly common detector technology used to measure neutrino interactions~\cite{DUNE:2020ypp,MicroBooNE:2016pwy,Machado:2019oxb,ArgoNeuT:2011bms}.}
{At energies of $\mathcal{O}(100)\si\MeV$ or greater, electrons and photons produced in such interactions will undergo a chain of bremsstrahlung radiation ($e^\pm \to e^\pm \gamma$) and pair production $\gamma \to e^+ e^-$) known as an ``electromagnetic shower''.}
Typical algorithms for the energy reconstruction of {such} electromagentic showers in LArTPCs have two main steps~\cite{MicroBooNE:2016dpb,MicroBooNE:2017xvs,MicroBooNE:2018kka,Qian:2018qbv}.
First, pixels are identified in the anode planes that measured charge associated with the shower of interest.
{For example,} this can be done by finding a cone that contains the shower and computing the total charge within the cone, excluding pixels with a low probability of being related to the shower~\cite{MicroBooNE:2021nss}.
Second, the total charge is linearly mapped to the shower energy.
{These traditional clustering-based linear algorithms are} fast, straightforward, and accurate for most LArTPC shower events. \\

However, when a LArTPC detector has sections of unresponsive wires, a subset of a shower's charge will pass undetected and the linear algorithm will underpredict the shower's energy.
This effect is more likely in higher energy showers, since their charge is spread over a larger area and has a higher chance of passing through an region of unresponsive wires.
In this case, the simple linear mapping to shower energy has difficulty correcting for the missing charge simultaneously across events of all energies, and events are often mis-reconstructed to lower energies. \\

A convolutional neural network (CNN) trained on a dataset with a substantial proportion of unresponsive wires could learn to reconstruct the energy of events with high fractions of missing charge.
CNNs train kernels to recognize patterns within an image, and thus efficiently use their parameters to extract information from very large image inputs. 
Visual features such as the shower size, radial extent, and longitudinal structure contain information about the shower's energy, so neural networks trained to recognize these patterns can {potentially} correct for the nonlinear mapping between charge and shower energy.  
Additionally, the position of unresponsive wires can be directly provided as input to the neural network to enable even more accurate reconstruction. \\

This study investigates the energy reconstruction performance of CNN models with varied sizes, architectures, training parameters, and inputs. We then compare the top CNN models' performance to that of traditional energy reconstruction methods under different detector conditions. We show that the CNN method is able to address inefficiencies caused by unresponsive wires in LArTPCs and reconstruct a larger fraction of imperfect events.

\section{Methods}
\label{sec:methods}
\subsection{Dataset}
\label{section:dataset}

We used the PILArNet public simulation pipeline~\cite{pilarnet2020} to generate simulated LArTPC images.
We assume a detector setup that includes three parallel anode planes with wires, spaced 3mm apart, oriented at $\SI{+60}\degree$ (U), $\SI{-60}\degree$ (V), and $\SI{0}\degree$ (Y) with respect to the zenith.
The PILArNet simulation process first instantiates a collection of particles at a vertex with randomly distributed kinetic energies, then tracks their charge deposition through the detector volume using GEANT4~\cite{GEANT4:2002zbu}.
The 3D charge information is voxelized into $3\si\mm$ length cubes and recorded, along with summary information on every particle generated, in a sparse data file.
Each pixel in Pilarnet images has an intensity value intended to represent the charge deposited in that region of the TPC. Hereafter we refer to this intensity value as the charge “Q”.
The PILArNet dataset is described in full in Ref.~\cite{Adams:2020vlj}. \\

The PILArNet procedure produces 3D voxel data for monodirectional electron showers. 
However, a LArTPC reads out charge information projected onto the anode planes through the drift process, and we are interested in showers of all orientations.
To adapt the PILArNet dataset to our needs, we rotate and project each shower event onto the anode planes to get a collection of isotropically oriented LArTPC events as they would be readout by the U, V, and Y anode planes.
Every event is assigned a random rotation, which is composed with a $\SI{+60}\degree$, $\SI{-60}\degree$, or $\SI{0}\degree$ degree rotation in the X-Y plane to get the three anode plane views.
For each rotation applied to an event, the charge in every input voxel is distributed to adjacent voxels in the rotated voxel grid with a weight inversely proportional to the euclidean distance between voxel centers.
This method is simple and preserves the total charge in the image.
Since the voxels are small, the particular choice of reapportionment metric is not significant.
Finally, the two dimensional anode plane images are created by summing the charge from voxels along the axis perpendicular to the wire planes.\\

\begin{figure}
\subfloat{\includegraphics[width=\textwidth]{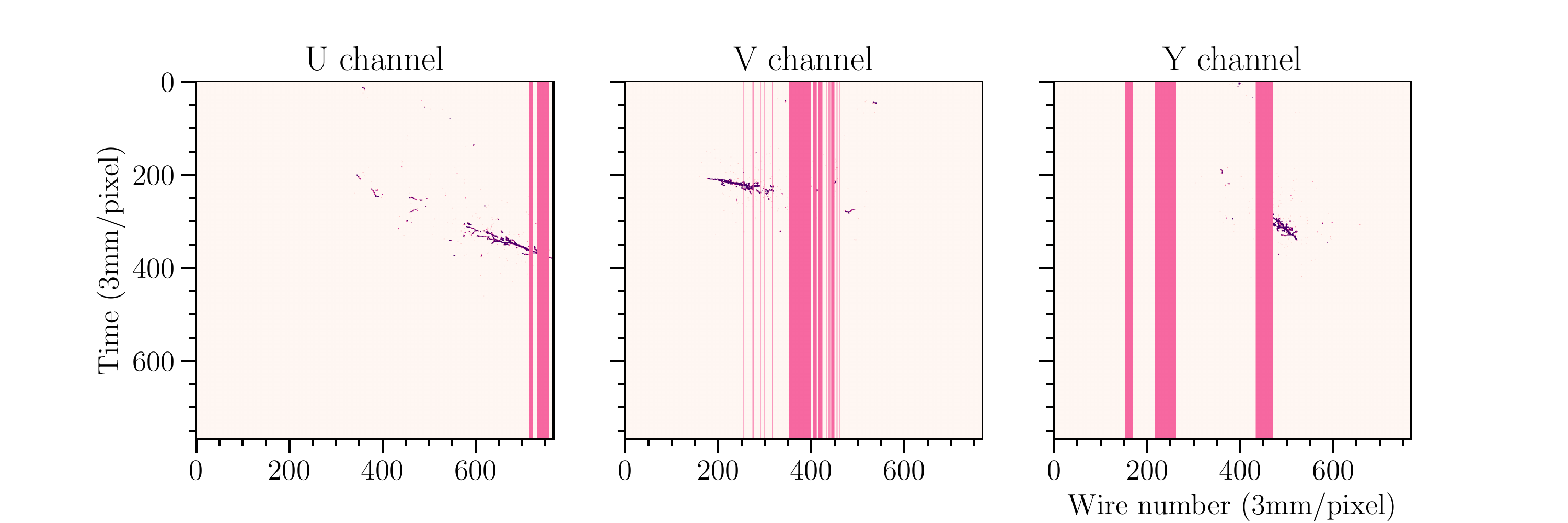}}
\caption{An example electron shower event (dark purple) from our testing dataset.
The locations of unresponsive wires are marked by superimposed pink stripes.}
\label{fig:example_img}
\end{figure}

Our resulting simulated LArTPC images are comprised of $600~\si\kilo$ single-electron shower events of energies uniformly distributed between $0$ and $\SI{1000}\MeV$ with a uniform angular distribution.
Uniform distributions were chosen to minimize any energy based or angular bias of the predictors.
We divided our dataset into $400~\si\kilo$ training events and $200~\si\kilo$ validation events.
Each event in the samples contains three base $768\times768$ images, which are the U, V, and Y plane readouts.
An example event image is displayed in Figure \ref{fig:example_img}.\\

We further augment our base dataset by adding unresponsive wires.
First, a set of unresponsive wire chunks is placed randomly within the full extent of the anode planes with $20~\si\kilo$, $20~\si\kilo$, and $21.5~\si\kilo$ wires composing each plane (these dimensions match the planned size of the DUNE far detector \cite{DUNE:2020txw}).
The unresponsive wire chunk lengths were randomly sampled to be consistent with typical ASIC sizes in a LArTPC: most chunks were less than 20 wires long, and all were at most 70.
The total fraction of unresponsive wires was fixed to $\SI{10}\percent$ for each plane, which is the magnitude reported by a 2017 MicroBooNE analysis of noise features in their LArTPC \cite{MicroBooNE:2017qiu}.
We then assigned each event a location within the detector volume and zeroed out all charge corresponding to unresponsive wires. \\

\subsection{Clustering-based Linear Algorithm}
\label{section:clustering}

The typical clustering algorithm approach to energy reconstruction identifies the shower of interest, and linearly calibrates the charge associated with the shower to a shower energy.
Our simulated dataset contains only single electron shower events, with no background.
Thus in our implementation of the traditional energy reconstruction algorithm, we linearly map the total charge in the Y collection plane image to a shower energy, bypassing the clustering step. \\

To find our linear mapping parameters, we first separate our event sample into 50 bins in total-Y-plane-charge, each $\SI{20}Q$ wide.
Within each total-charge bin the distribution of true shower energies is fit to a Gaussian distribution.
Any bins for which the Gaussian fit is sufficiently imprecise are skipped: that is, the bins are skipped if the fit's uncertainty on a parameter is greater than ten times the parameter's value.
The remaining bins are interpreted as observations of the true shower energy at specific total charges with corresponding errors.
These data points are then used as input to a least squares fit of the linear mapping between the binned shower charge and true shower energy.
On datasets with unresponsive wires, we found this procedure corrected for bias in all energy bins over $\SI{60}\MeV$, but did not do so in bins below.

\subsection{Convolutional Neural Networks}
\label{section:CNNs}

Different neural network architectures can be better suited for different problems. For this study, we focused on two neural network architectures: the residual network~\cite{He2016DeepRL} and the inception network~\cite{2014arXiv1409.4842S}. Each has found success in previous LArTPC deep learning projects~\cite{Baldi:2018qhe,MicroBooNE:2020yze} \\

The residual network (ResNet~\cite{He2016DeepRL}), introduced in 2015, was designed to allow deep layers in the network to learn small adjustments to the final output easily.
Its central structure is the residual block, a network layer whose output is a sum of its input and a filtered ``residual.'' The original authors suggested that because residual blocks make it easy for deep layers to implement an identity map, very deep ResNets can converge more quickly to more accurate solutions. \\

The first version of the inception network \cite{2014arXiv1409.4842S} was published in 2014 and designed as a wider, rather than deeper, network.
For the purposes of this paper we have labeled this architecture InceptNet.
Its fundamental structure is the inception block, which combines the results of convolutional filters of various sizes into one output.
Inception blocks are therefore suited to describing features of different sizes simultaneously.
The original authors suggested that such hybrid blocks could approximate a sparse network structure by dense component calculations.
Thus, this architecture might work well for LArTPC images, which are sparse by nature.
A variation of the inception network was successfuly implemented for energy reconstruction by the $\text{NO}\nu\text{A}$ collaboration in 2019 \cite{Baldi:2018qhe}. \\

The original authors of the inception network have since released three updated versions.
These updates employ new structural efficiencies to express the same operations in fewer calculations, and thus can form even deeper networks.
However, these architectures quickly grow too large for our purposes, so for this paper we focus solely on the original version. \\

\begin{figure}
\caption{The basic structure of a ResNet consists of a starting gate, a stack of residual layers, and then a fully-connected layer which decodes vector output into a single energy prediction \textit{(left)}. Each residual layer is formed out of stacks of the basic residual block (\textit{right}) which transforms its input by adding a small convolutional correction.  }
\subfloat{\includegraphics[width=\textwidth]{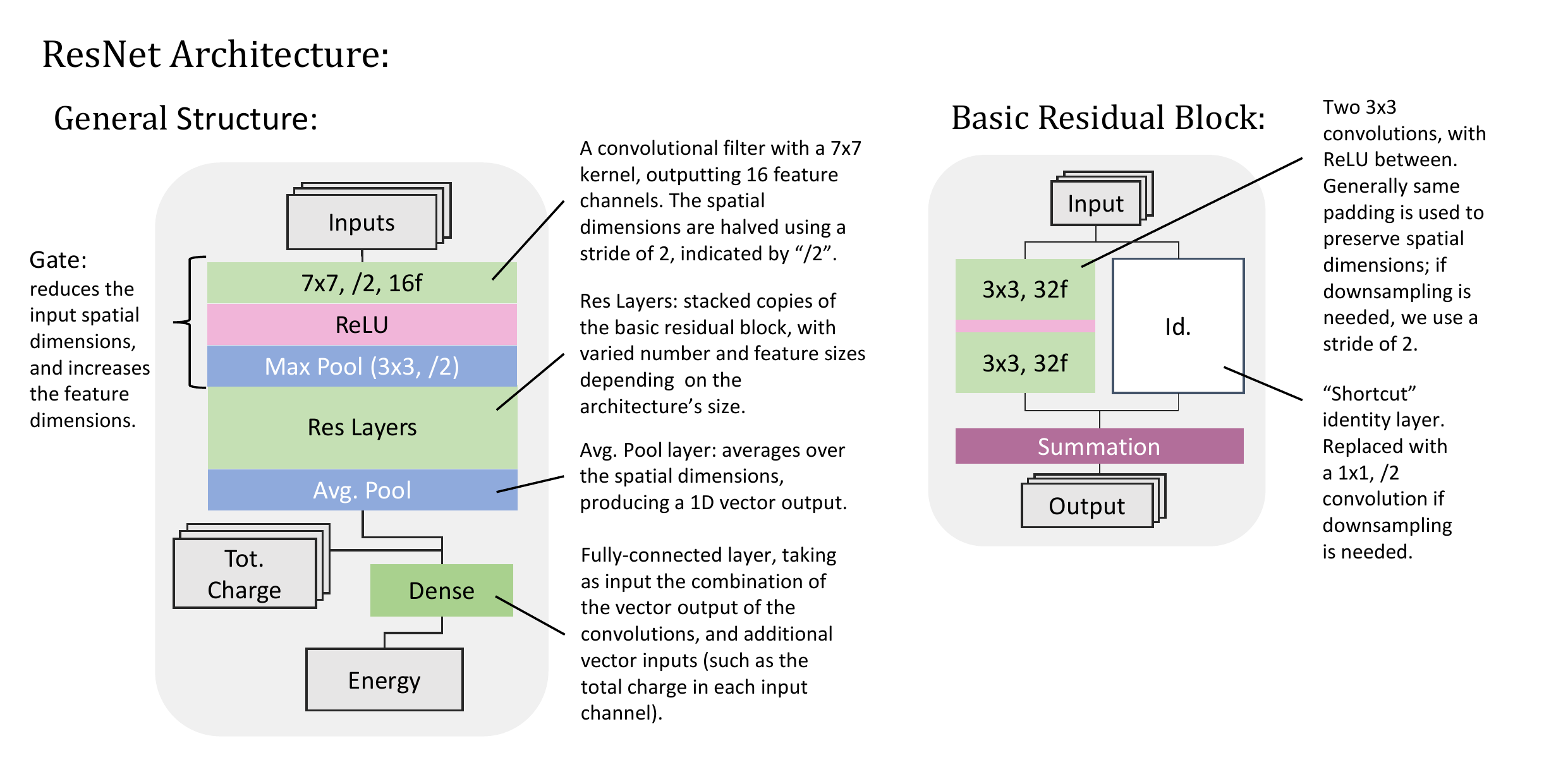}}
\label{fig:arch_res_general}
\end{figure}

\begin{figure}
\caption{The basic structure of our InceptionNet is similar to that of the ResNet, but includes a slightly larger gate (\textit{left}). Inception layers consist of stacks of the basic inception block, which concatenates the output of convolutional filters of many different sizes (\textit{right}). Concatenation is possible since all convolutional filters use the same padding, and thus preserve spatial input dimensions.}
\subfloat{\includegraphics[width=\textwidth]{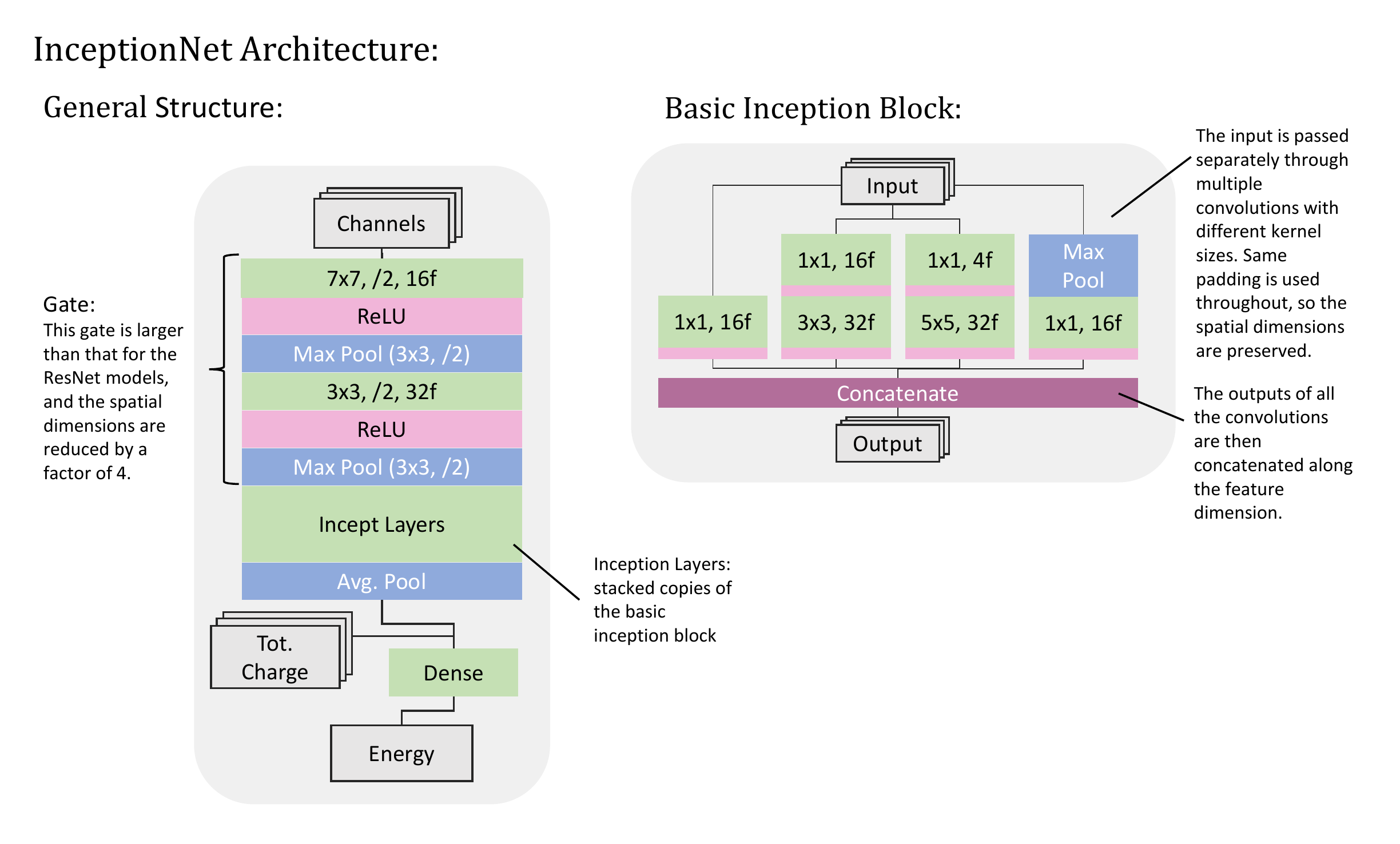}}
\label{fig:arch_incept_general}
\end{figure}

We implemented both the ResNet and InceptNet architectures according to the same general structure as the original works. 
These are, respectively, depicted schematically in Figures~\ref{fig:arch_res_general} and~\ref{fig:arch_incept_general}. 
Each network has an initial gate, which scales down the image spatial dimensions and increases the feature dimensions. 
The gate is then followed by a stack of convolutional layers, which constitute the bulk of the network. 
For each architecture, we built models of four different sizes, ``Small,'' ``Medium,'' ``Large,'' and ``Huge,'' by increasing the depth of this stack.
Each network size has roughly an order of magnitude more trainable parameters than the previous.
We scaled architectures up according to two main principles: first, by stacking on additional layers of the architecture's basic block, and second, by adjusting the feature map sizes throughout the network so that the number of parameters in sequential layers increases gradually, the input image dimensions decrease gradually, and the feature dimension increases.
The exact implementation of each network layer is most likely not significant~\cite{2014arXiv1409.4842S}.
Details of the four network sizes are given in Appendix~\ref{app:archs}, and Figures~\ref{fig:arch_res_sizes} and~\ref{fig:arch_incept_sizes}.
The quantitative properties of each model size, including the exact numbers of parameters, are listed in Table \ref{table:model_params} and are discussed later in this work.
The output of the convolutional layers is collapsed into a vector by an average pool layer.
This vector can then optionally be combined with additional vector inputs, and is finally passed through a dense fully-connected layer, which outputs the predicted energy.  \\

The networks take as input a batch of layered images and vectors of additional information.
The primary inputs were the U, V, Y output planes of the LArTPC detector, and a three-component vector containing the total recorded charge on each plane.
We also investigated whether the the networks could utilize information about unresponsive wire placement.
In one set of tests, we fed in images with six channels: three base U, V, Y $768 \times 768$ pixel images, and three $768 \times 768$ images indicating the locations of dead wires for the corresponding planes. \\

CNNs are trained using stochastic gradient descent to iteratively adjust the network's weights such that its performance is improved with respect to some loss function~\cite{bottou-98x}.
To reduce memory costs associated with large image-based datasets, during optimization the training set is sampled in smaller batches.
The model's weights are thus adjusted in a series of less-confident steps, rather than one sure leap.
The weight adjustment step for each batch can be further modified to incorporate information about the previous adjustment: the parameters at batch $t+1$ are related to those at batch $t$ by $p_{t+1} = p_{t} - m \cdot u_t + \ell \cdot g_{t+1} $, where $g_{t+1}$ is the gradient for batch $t+1$, $u_t= p_{t-1} - p_t$ is the adjustment factor for batch $t$, and $\ell, m$ are fixed hyperparameters called the learning rate and momentum respectively. 
We tested different loss functions to explore how they would affect the final network performance at different energies. \\ 

We implemented and trained our networks using PyTorch's software library~\cite{NEURIPS2019_9015}, and we optimized our weights using PyTorch's stochastic gradient descent algorithm with an initial learning rate of $\ell = 0.1$ and a momentum factor $m = 0.5$.
Since we expected training to converge on some local optimal weight configuration, we reduced our learning rate when performance improvements plateaued: if for ten epochs the total loss did not decrease below $0.9999$ times the best value, we reduced the learning rate by a factor of 10. \\

Training was divided into epochs.
In each epoch, the network learned from a randomized subset of $160~\si\kilo$ images from the $400~\si\kilo$ event training dataset, and validated its results on a randomized subset of $6~\si\kilo$ images from the $200~\si\kilo$ event testing dataset.
Due to time and GPU memory constraints, we trained all models for 100 epochs total.
We then tested model performance using the trained parameters from the 100\ts{th} epoch, which were generally the most refined.
We expect all the models would see additional small performance gains if trained further, especially those using the Residual architecture.\\

\section{Experiments and Results}
\label{sec:experiments_results}

In this section, we compare the performances of the different networks described in the previous section. 
In Section~\ref{section:size}, we examine the performance as a function of network size. 
Next, in Section~\ref{section:loss}, we compare networks trained using three different loss functions. 
In Section~\ref{section:inputs}, we evaluate the network performance with and without including information on unresponsive wires. 
Finally, in Section~\ref{sec:linear}, we compare the performance of the neural networks to that of a linear reconstruction algorithm.
See Appendix~\ref{app:details} for more details on the studies described in this section.

\subsection{Size Comparison}
\label{section:size}

The modular designs of the ResNet and InceptNet architectures imply that we must choose the size of the network.
In general, an increased number of training parameters will allow the network to capture more complex structure.
However, larger network sizes come with a substantial increase in the training time and memory consumption, and so it is worth investigating the reconstruction performance as a function of network size.
Since LArTPC images are relatively sparse compared to other datasets like photographic images, we expect that increasing the network size should have diminishing returns~\cite{Baldi:2018qhe}. \\

We tested models of four different sizes, ``Small'', ``Medium'', ``Large'', and ``Huge'', as described in \ref{section:CNNs}. 
Table~\ref{table:model_params} contains the quantitative properties of each size model.
The number of parameters scaled linearly with the space in memory required to store the model, and non-linearly with the training and evaluation time needed per image. \\

\begin{table}[h]
\centering
\begin{tabular}{l l l l l}
\toprule
\makecell[tl]{Architecture} & \makecell[tl]{Size} & \makecell[tl]{Number of\\Parameters [M]} & \makecell[tl]{Training\\Time / Image [ms]} & \makecell[tl]{Validation\\Time / Image [ms]} \\
\midrule
InceptNet & Small     & 0.088   & $2.09 \pm 0.05$ & $0.60 \pm 0.003$ \\
InceptNet & Medium & 0.915   & $2.71 \pm 0.49$ & $0.73 \pm 0.002$ \\
InceptNet & Large    & 5.798   & $4.53 \pm 0.46$ & $1.18 \pm 0.002$ \\
InceptNet & Huge     & 21.361 & $8.88 \pm 0.68$ & $2.25 \pm 0.004$ \\
ResNet & Small     & 0.082       & $2.50 \pm 0.34$ & $0.66 \pm 0.002$ \\
ResNet & Medium & 0.705       & $3.51 \pm 0.38$ & $0.85 \pm 0.0001$ \\
ResNet & Large    & 4.378       & $4.61 \pm 0.04$ & $1.07 \pm 0.0001$ \\
ResNet & Huge     & 23.786     & $6.12 \pm 0.80$ & $1.44 \pm 0.003$ \\
\bottomrule
\end{tabular}
\caption{Quantitative properties of different-sized models. Each increase in model size corresponded to an order of magnitude increase in the number of parameters.}
\label{table:model_params}
\end{table}

\begin{figure}[h]
\subfloat{\includegraphics[width=0.5\textwidth]{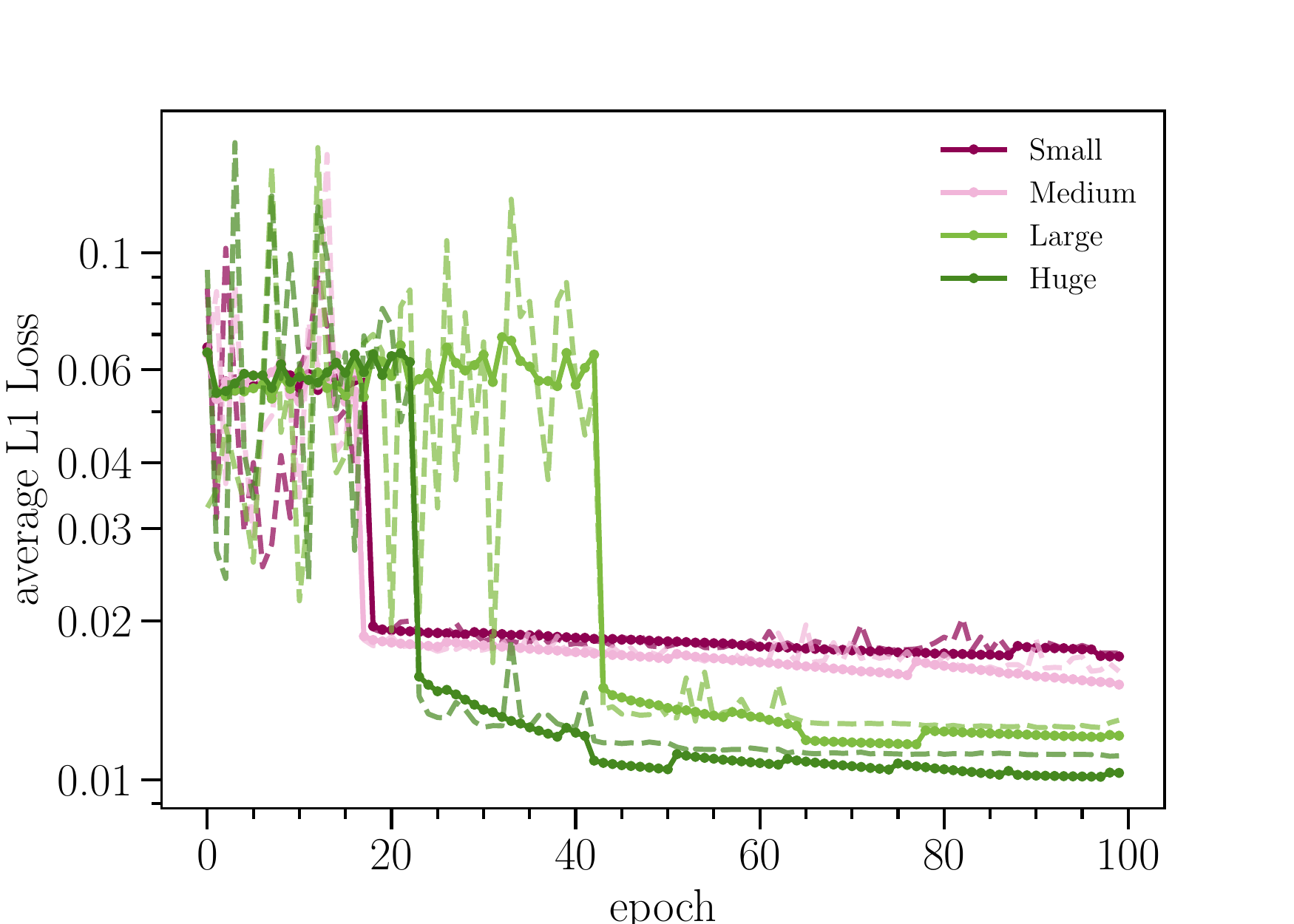}}
\subfloat{\includegraphics[width=0.5\textwidth]{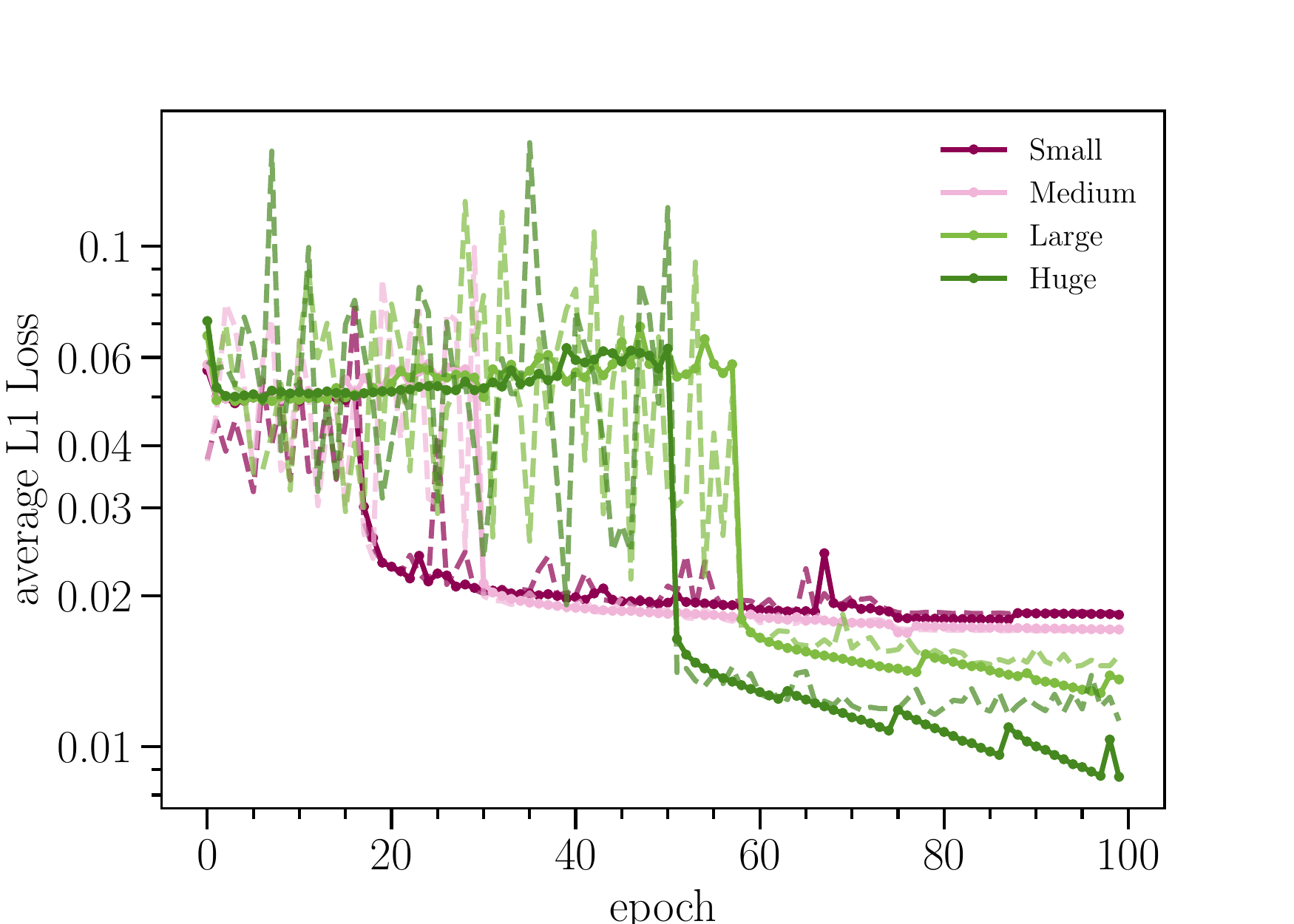}}
\caption{The average training (solid) and testing (dashed) L1 loss per epoch, for the different sized InceptNet (\textit{left}) and ResNet (\textit{right}) models. 10-epoch periodic dips in the training loss coincide with scheduled reductions of the learning rate.}
\label{fig:size_loss}
\end{figure}

All models were trained on training and evaluation sample datasets of $400~\si\kilo$ and $200~\si\kilo$ images each, using the setup described above with a linear (L1) loss function.
In this comparison, we used the datasets with unresponsive wires for both training and testing.
In order to use the memory of our GPUs efficiently, we trained the huge model with a batch size of 64 images, and all the other sizes 128.
The loss curve for every model, depicted in Figure \ref{fig:size_loss}, shows that substantial learning occurs by epoch 40.
After this point, the loss curves show slight improvements at about every ten epochs, when the learning rate decreased. 
We expect training for additional epochs could yield small additional performance improvements.\\

As one would expect, the uncertainty on the reconstructed energy decreases as the network size increases. For the InceptNet Small, Medium, Large, and Huge models, the top $\SI{68}\percent$ of events have a reconstructed energy within $\SI{3.96}\percent$, $\SI{3.86}\percent$, $\SI{3.15}\percent$ and $\SI{2.79}\percent$ of the true value respectively, and for the corresponding ResNet models it is within $\SI{3.89}\percent$, $\SI{3.48}\percent$, $\SI{3.40}\percent$, and $\SI{2.56}\percent$. \\

\begin{figure}[h]
\subfloat{\includegraphics[width=0.48\textwidth]{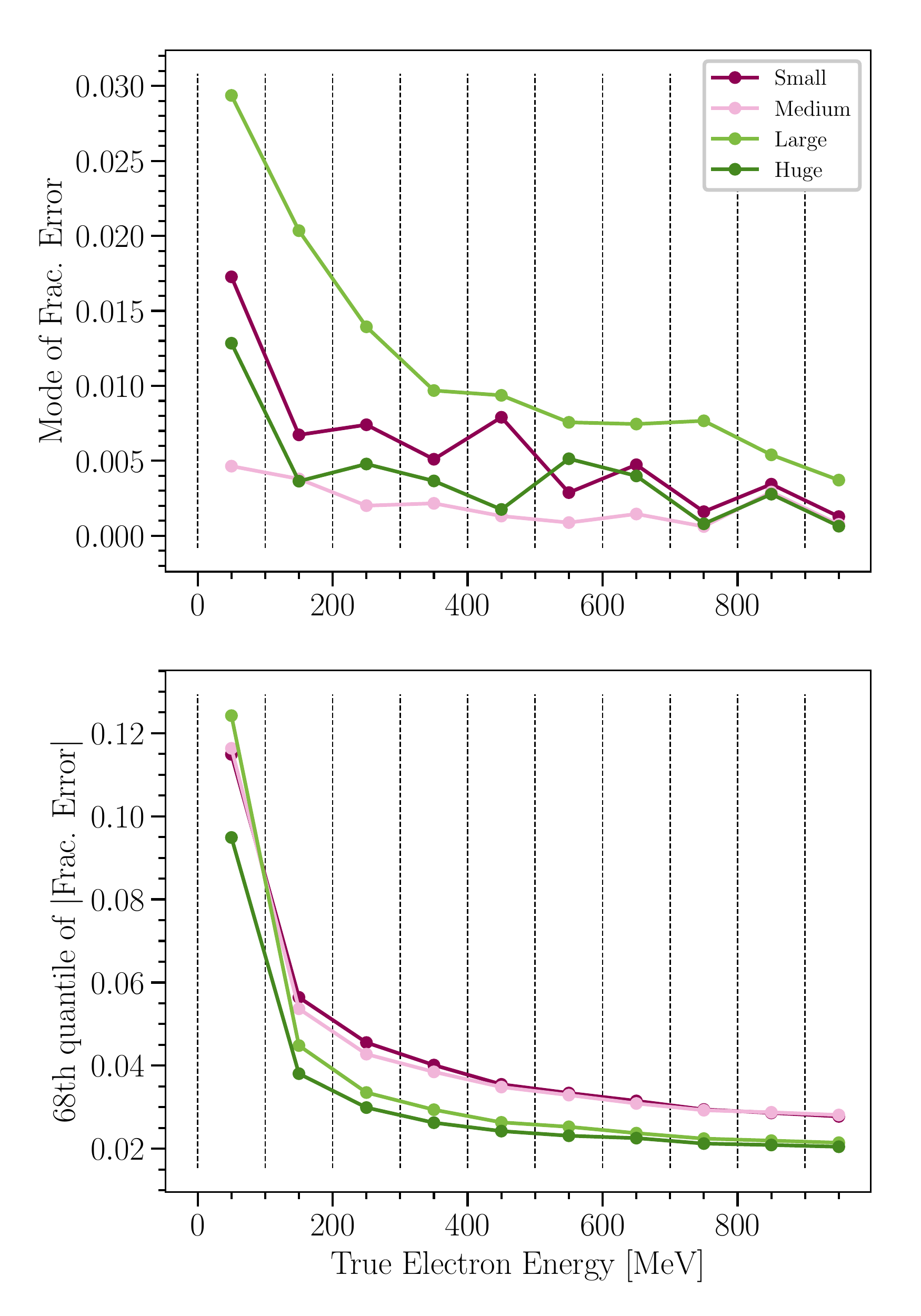}}
\subfloat{\includegraphics[width=0.48\textwidth]{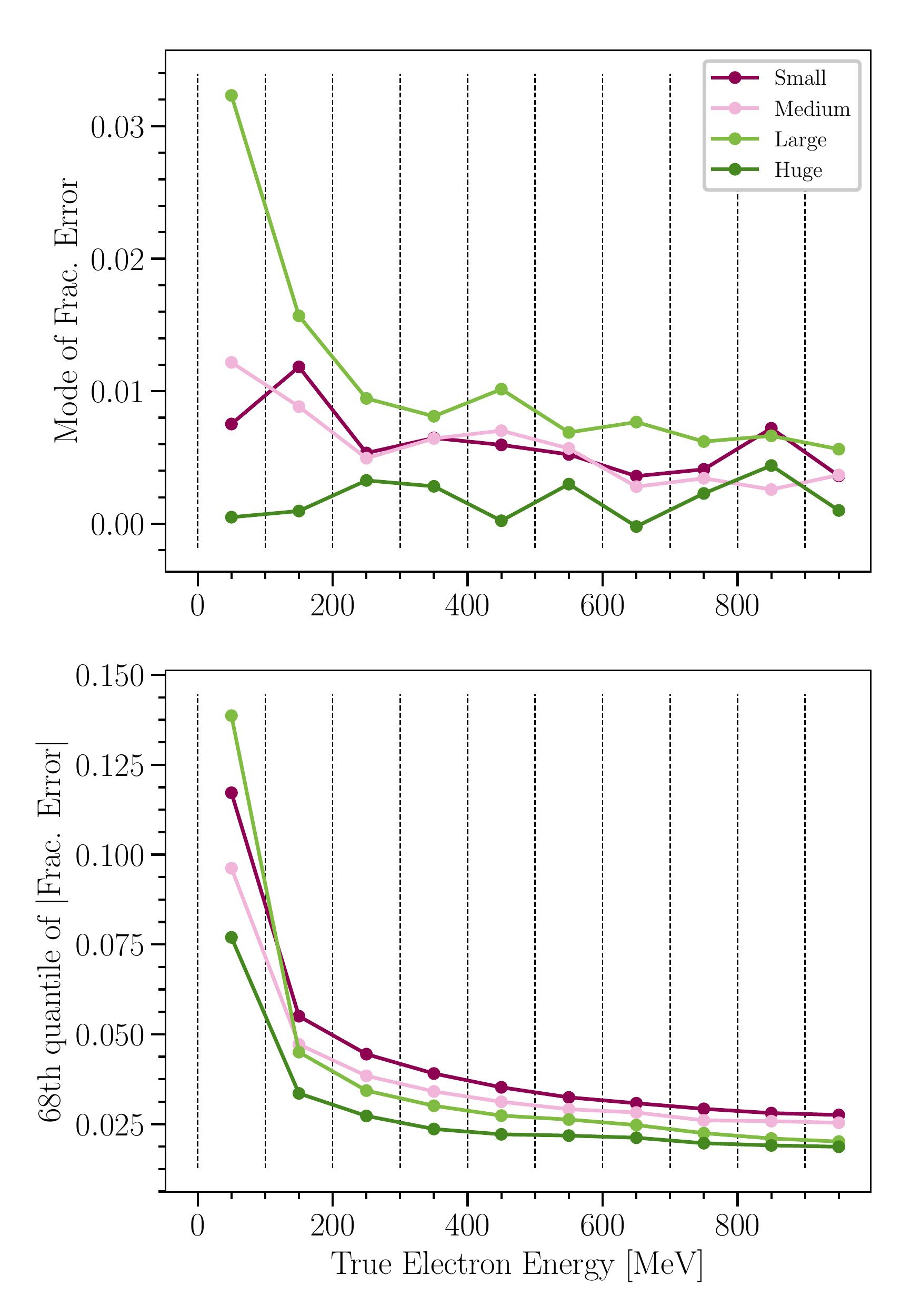}}
\caption{The mode and absolute 68th quantiles of the fractional reconstruction error, in each bucket of true shower energy, for the InceptNet models \textit{(left)} and the ResNet models \textit{(right)}.}
\label{fig:size_mpvq}
\end{figure}

Figure~\ref{fig:size_mpvq} depicts the performance of each model size as a function of the true shower energy.
In the upper plots, we see that the modes of the fractional error distributions are stable and close to zero across all energies.
Both the InceptNet and ResNet large models show an at least $\SI{1}\percent$ overprediction bias at all energies.
In the lower plots, we see that the widths of the distributions improve across all energies with increasing model sizes.
All models reconstruct the lowest energy bin, $\SI[parse-numbers=false]{0-100}\MeV$ much less precisely than higher energy bins. \\

The Large networks show a visible overprediction bias compared to the other sized networks across all energy bins, and a slightly larger overall overprediction bias (0.7\%, as opposed to 0.5\% or less). We did expect over prediction to be a more likely outcome than under, since for a large subset of events, the models need to compensate for missing charge during reconstruction. Ideally, this compensation would be regulated by factors such as large chunks of discontinuity in the shower or the known presence of unresponsive wires, but it's likely that a flat correction is also used. If this were the case, then the error (mode and spread) would be worst in the lowest energy bin, which is what we see in Figure~\ref{fig:size_mpvq}. We think it is unlikely that this bias is a feature of solely the Large sized network. The size of this effect may be an accident of the final state parameters, and could be corrected after additional epochs of training. \\

The Huge models produced the narrowest, most sharply peaked distribution across the full range of energies.
However, Huge models are computationally very expensive.
Not only is the per-image training time $\SI[parse-numbers=false]{150-200}\percent$ that of the Large models, but also the quadrupled model size limits the maximum number of events that can be batched together, so the training procedure needs to cycle smaller sets of events on and off the GPU.
For other applications, the performance improvement may be worth the extra cost in time and memory - for this study, the large models offer excellent predictive power, with small variance in fractional error, for a reasonable cost.

\subsection{Loss Function Comparison}
\label{section:loss}

For the next study, we compared the performance of networks trained with a linear (L1), fractional (Frac.), and mean squared error (MSE) loss function.
A linear loss function weights every event equally, while a squared error loss function weights outlier reconstructions more heavily.
The fractional loss function weights errors in low energy events more heavily, and using it to train models could prioritize the performance in those events.
Fractional loss was also used successfully by the $\text{NO}\nu\text{A}$ collaboration~\cite{Baldi:2018qhe}.
For this comparison, we used the Large network size for both the Inception and Residual architectures.
We trained these models as described in section~\ref{section:CNNs}, with batch sizes of 128 events. \\

\begin{figure}[h]
\subfloat{\includegraphics[width=0.5\textwidth]{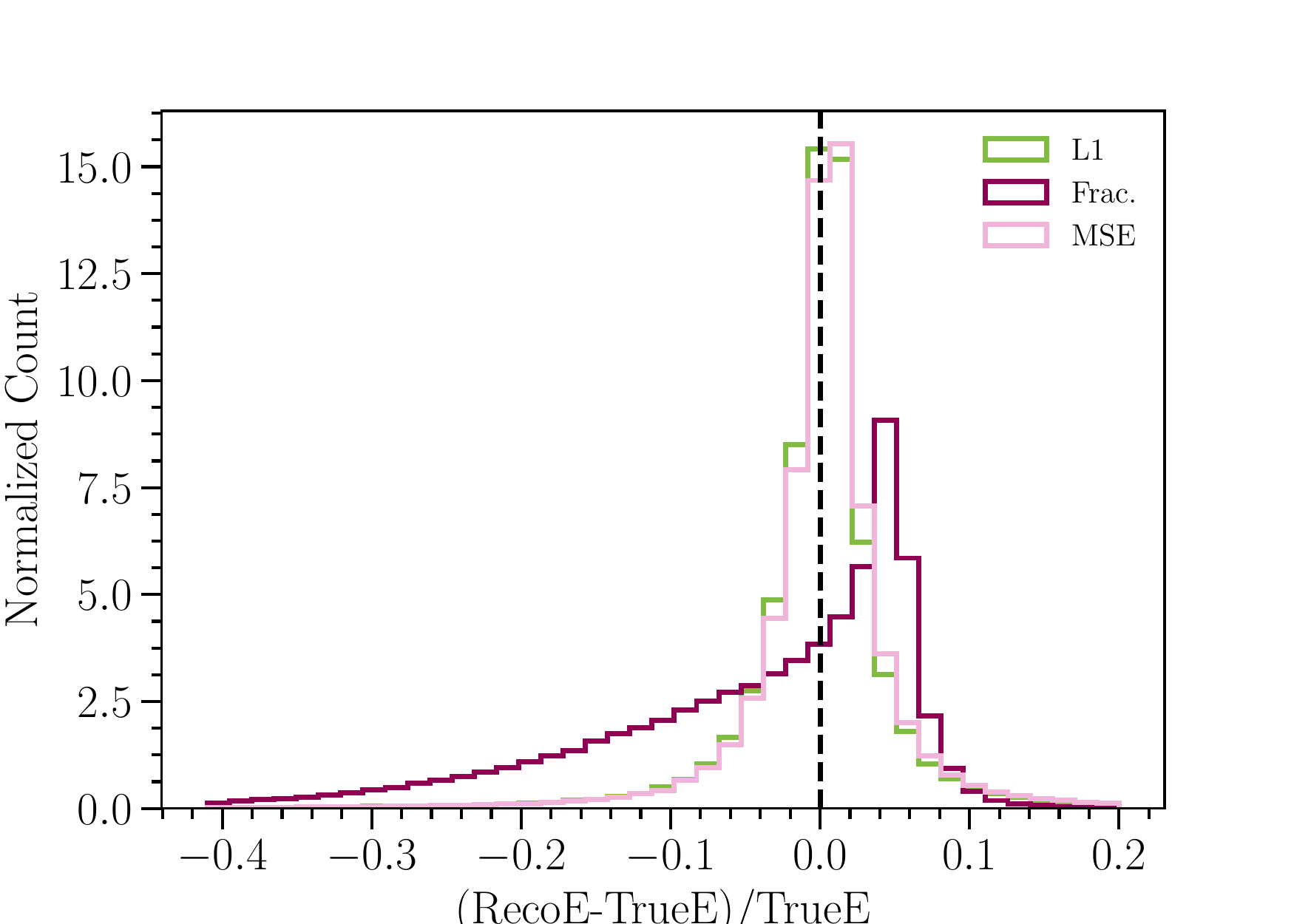}}
\subfloat{\includegraphics[width=0.5\textwidth]{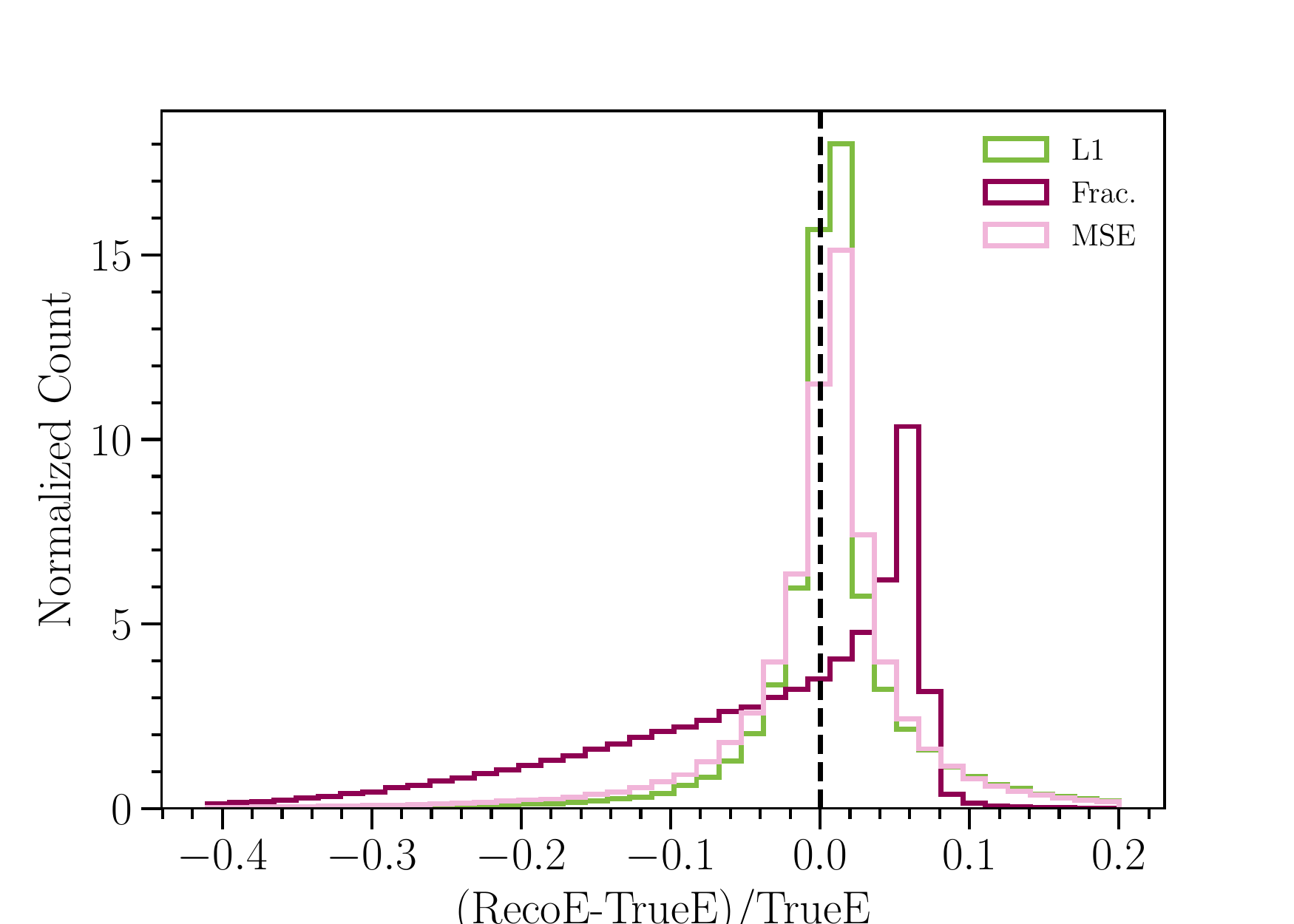}}
\caption{Histograms of the fractional error for the InceptNet models \textit{(left)} and the ResNet models \textit{(right)} trained with the three different loss functions. The models trained with the L1 and MSE loss perform comparably, but the fractional (Frac.) loss models underperform.}
\label{fig:loss_hist}
\end{figure}

\begin{figure}[h]
\subfloat{\includegraphics[width=0.48\textwidth]{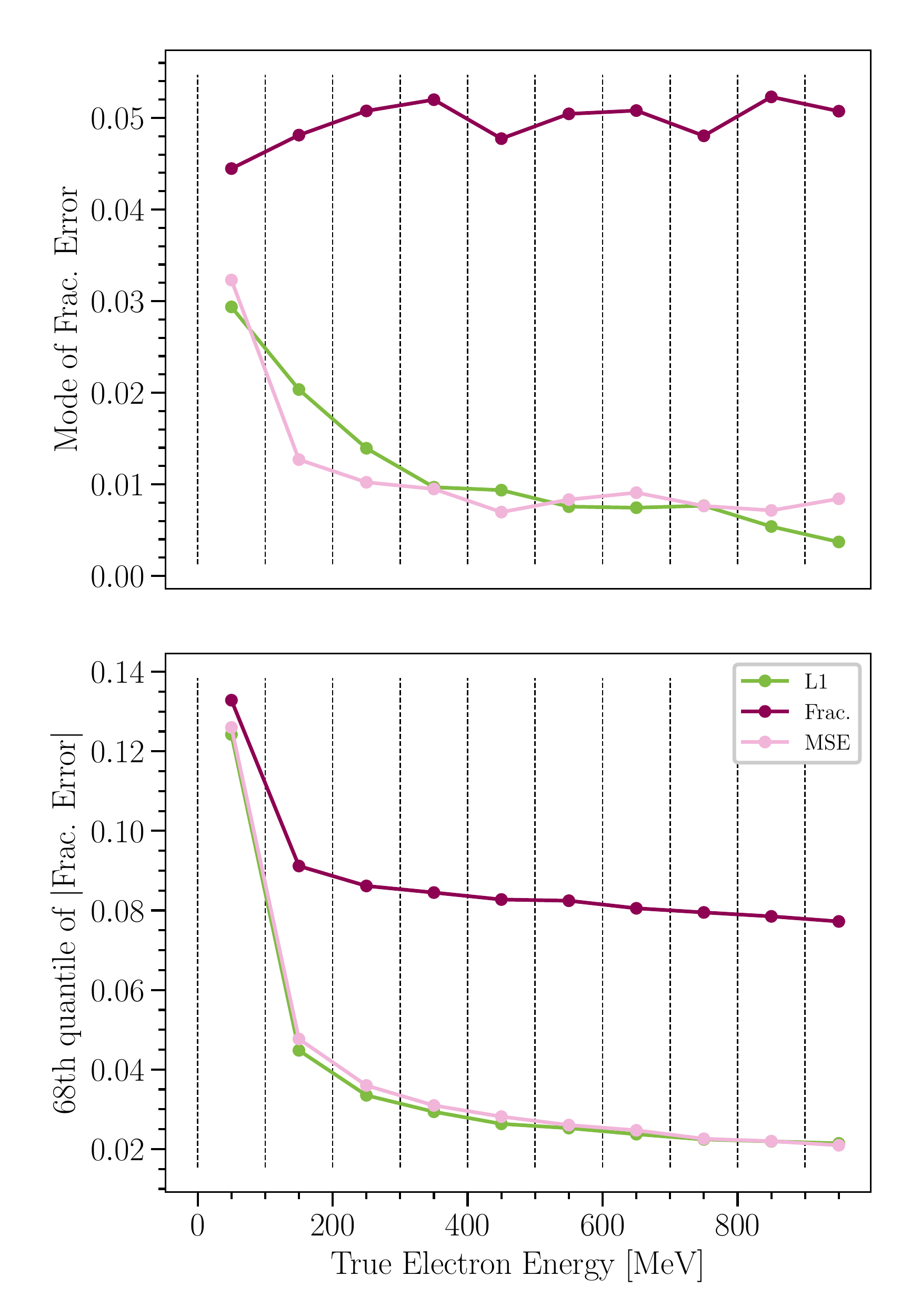}}
\subfloat{\includegraphics[width=0.48\textwidth]{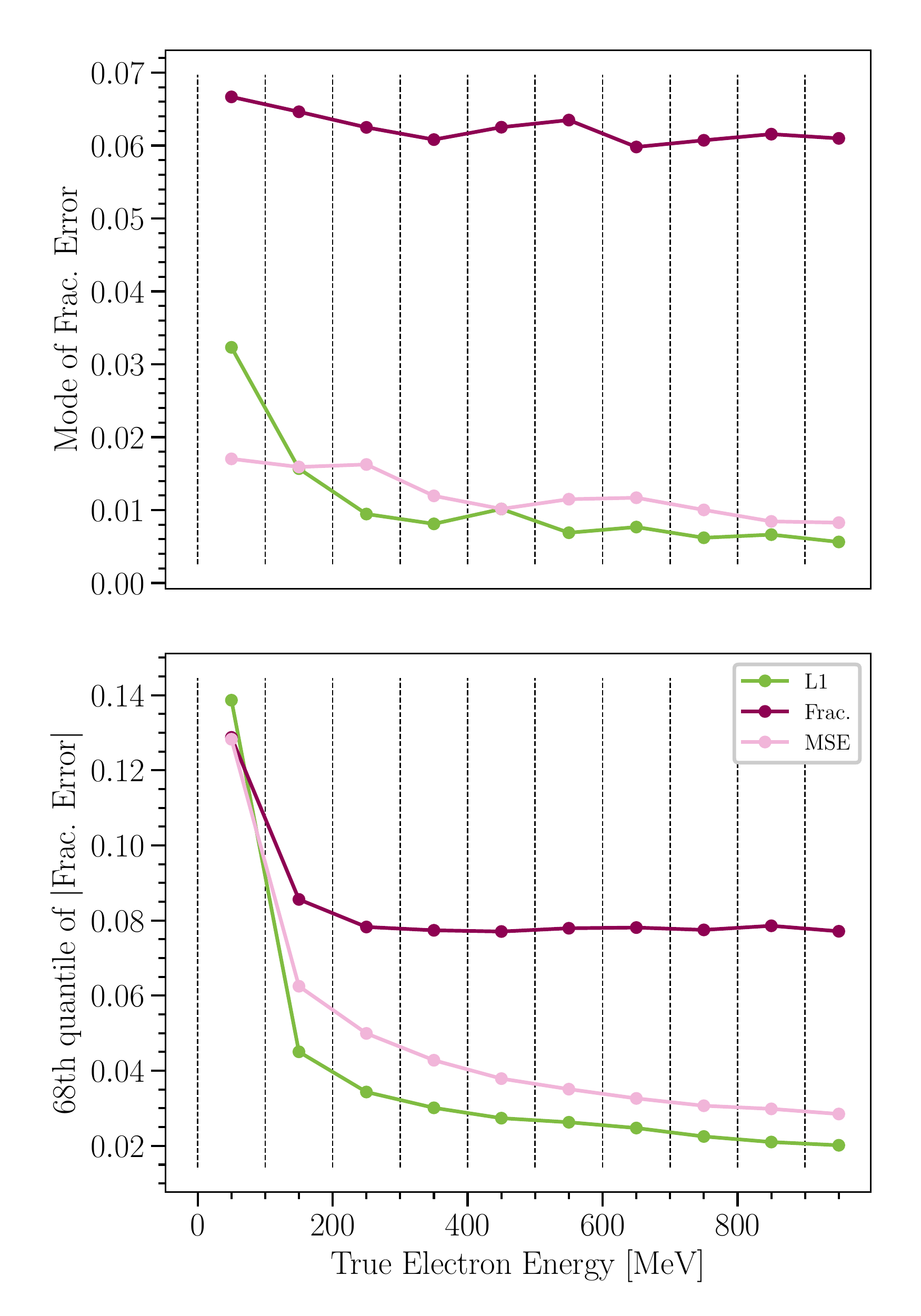}}
\caption{The mode and absolute $\SI{68}\percent$ quantiles of the fractional reconstruction error, in each bucket of true shower energy, for the InceptNet models \textit{(left)} and the ResNet models \textit{(right)} trained with different loss functions.}
\label{fig:loss_mpvq}
\end{figure}

The L1 and MSE loss functions perform similarly, but the L1 loss model has a narrower error distribution across all energy ranges. Therefore, we conclude that the L1 loss function is slightly more optimal than the MSE loss function.
Training with the fractional error loss produced skewed model predictions.
In Figure~\ref{fig:loss_hist} we plotted the fractional reconstruction error for each evaluated model.
The Frac. distribution is peaked at an overprediction of around $\SI{5}\percent$, and has a wide tail of underpredictions extending beyond $\SI{-40}\percent$.
The plots of the mode value and $\SI{68}\percent$ quantile as a function of energy in Figure~\ref{fig:loss_mpvq} further show that this $\SI{5}\percent$ overprediction bias and large distribution width extend throughout all true energy bins.
This behavior when using the fractional loss function is not entirely understood. It may be related to the specific interplay of the network architecture, loss function, and training data. \\

\subsection{Network Input Comparison}
\label{section:inputs}

We next investigate the importance of input information concerning unresponsive wires. In order to investigate the neural networks' ability to compensate for unresponsive wires in the detector, we trained three groups of models on three different sets of information.
The first group was trained on an ideal dataset with no unresponsive wires.
The second group was trained on a dataset with an average of $\SI{10}\percent$ unresponsive wires in each plane (URW dataset), and was given information on the locations of these wires.
We marked the locations of responsive or unresponsive wires on each plane with pixel values of zero or one respectively, and recorded these on a separate set of three planes per event which were given as additional input to the neural network.
The last group was trained on the same dataset as the second, but was not given any wire information. \\

We tried both the Inception and Residual architectures for this comparison, but used the results from sections~\ref{section:size} and~\ref{section:loss} to fix the model size to Large and the loss function to L1.
All models were trained on a set of $400~\si\kilo$ images and validated on a set of $200~\si\kilo$, with a batch size of 128, for a total of 100 epochs, as described in section~\ref{section:CNNs}. \\

\begin{figure}[h]
\subfloat{\includegraphics[width=0.5\textwidth]{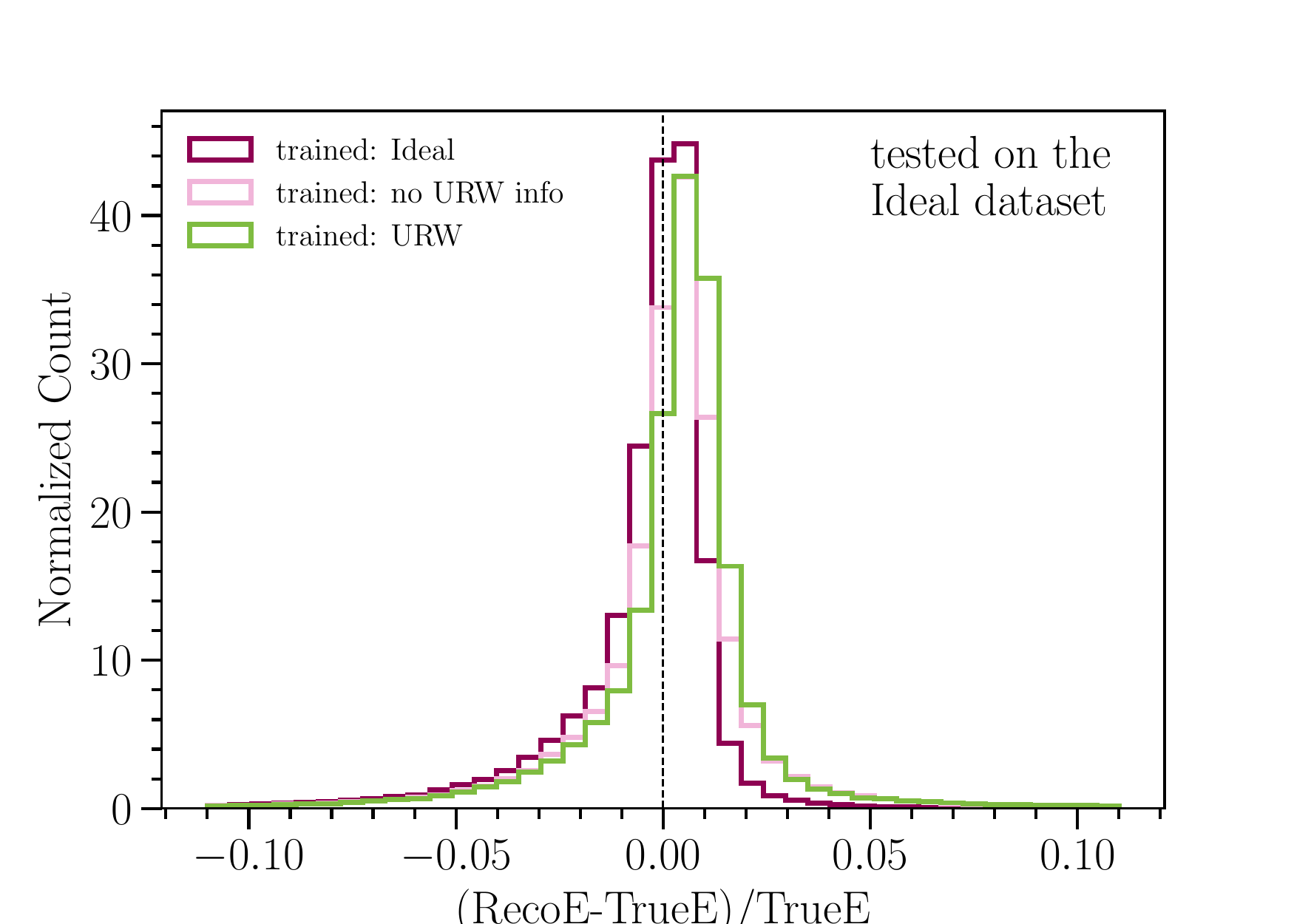}}
\subfloat{\includegraphics[width=0.5\textwidth]{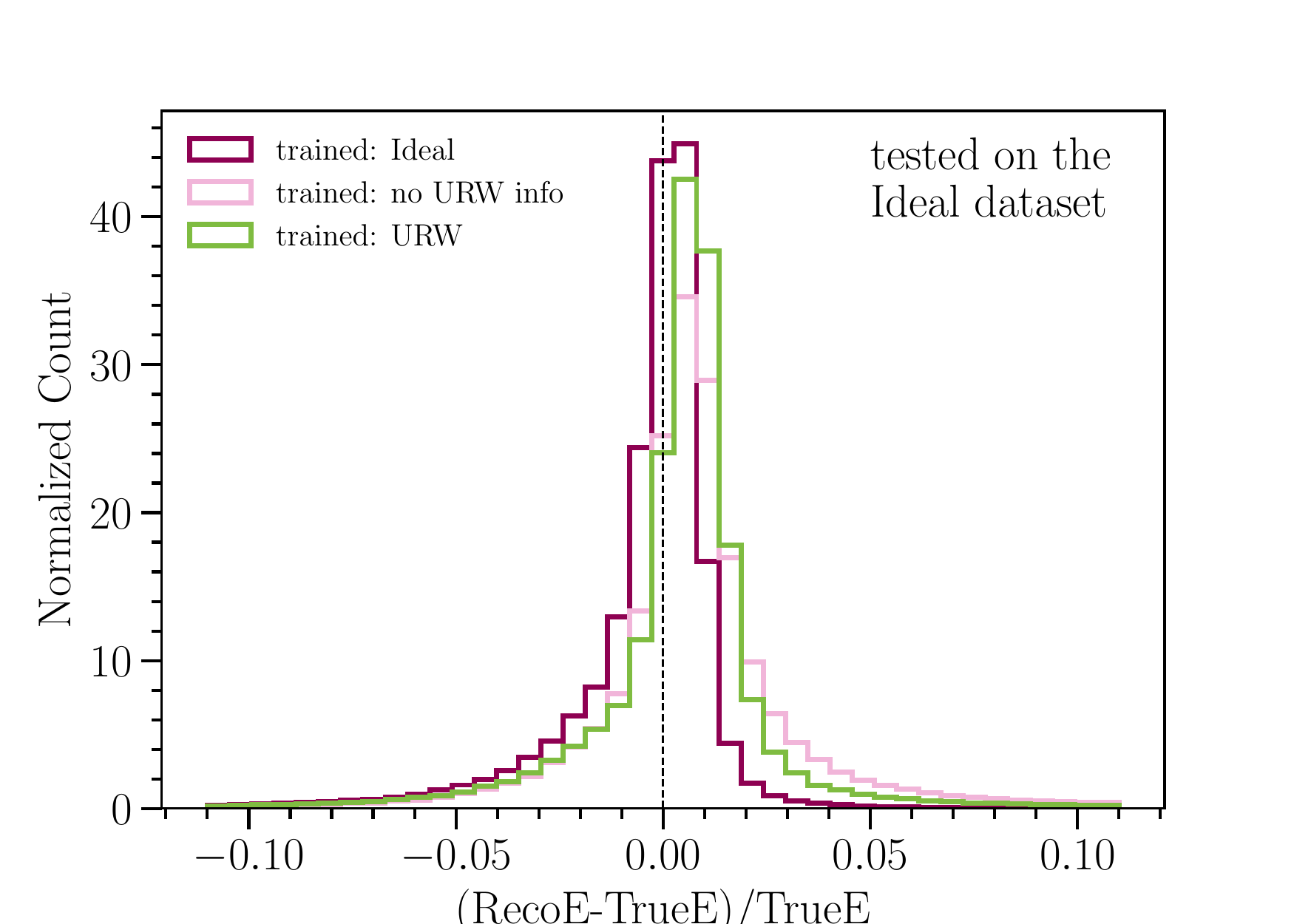}}
\caption{The fractional error of the InceptNet models (\textit{left}) and ResNet models (\textit{right}) trained on three different input datasets, and evaluated on the validation dataset with ideal, no unresponsive wire images.}
\label{fig:inputs_ideal}
\end{figure}

We then tested the performance of these three groups of models on $200~\si\kilo$ validation events in both the ideal dataset and on the $\SI{10}\percent$ unresponsive dataset.
Figure~\ref{fig:inputs_ideal} the fractional error distribution on the ideal dataset for the InceptNet and ResNet models of the three groups.
All the models output reconstructed energies which are within $\SI{2}\percent$ of the true values for greater than $\SI{68}\percent$ of all test events.
However, the models trained on the unresponsive wire datasets slightly overpredict the energy.
$\SI{49.8}\percent$ of the ideal InceptNet reconstructions are greater than the true energy, but $\SI{68.9}\percent$ ($\SI{62.3}\percent$) of the informed (uninformed) unresponsive wire InceptNet reconstructions are greater than the true energy.
Similarly, $\SI{49.9}\percent$ of the ideal ResNet reconstructions are greater than the true energy, but $\SI{71.2}\percent$ ($\SI{71.4}\percent$) of the informed (uninformed) unresponsive wire ResNet reconstructions respectively are greater than the true energy.\\

\begin{figure}[h]
\subfloat{\includegraphics[width=0.48\textwidth]{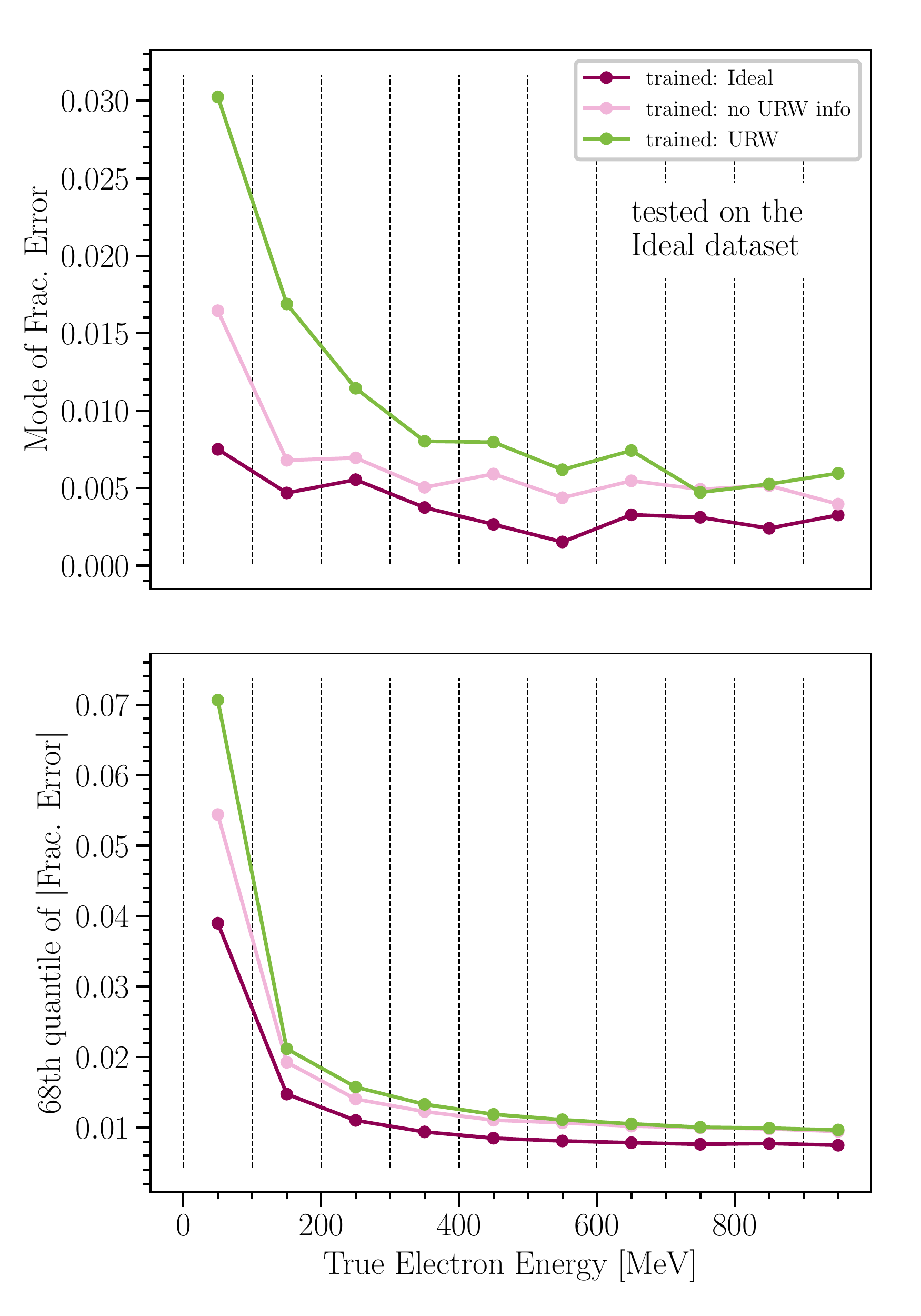}}
\subfloat{\includegraphics[width=0.48\textwidth]{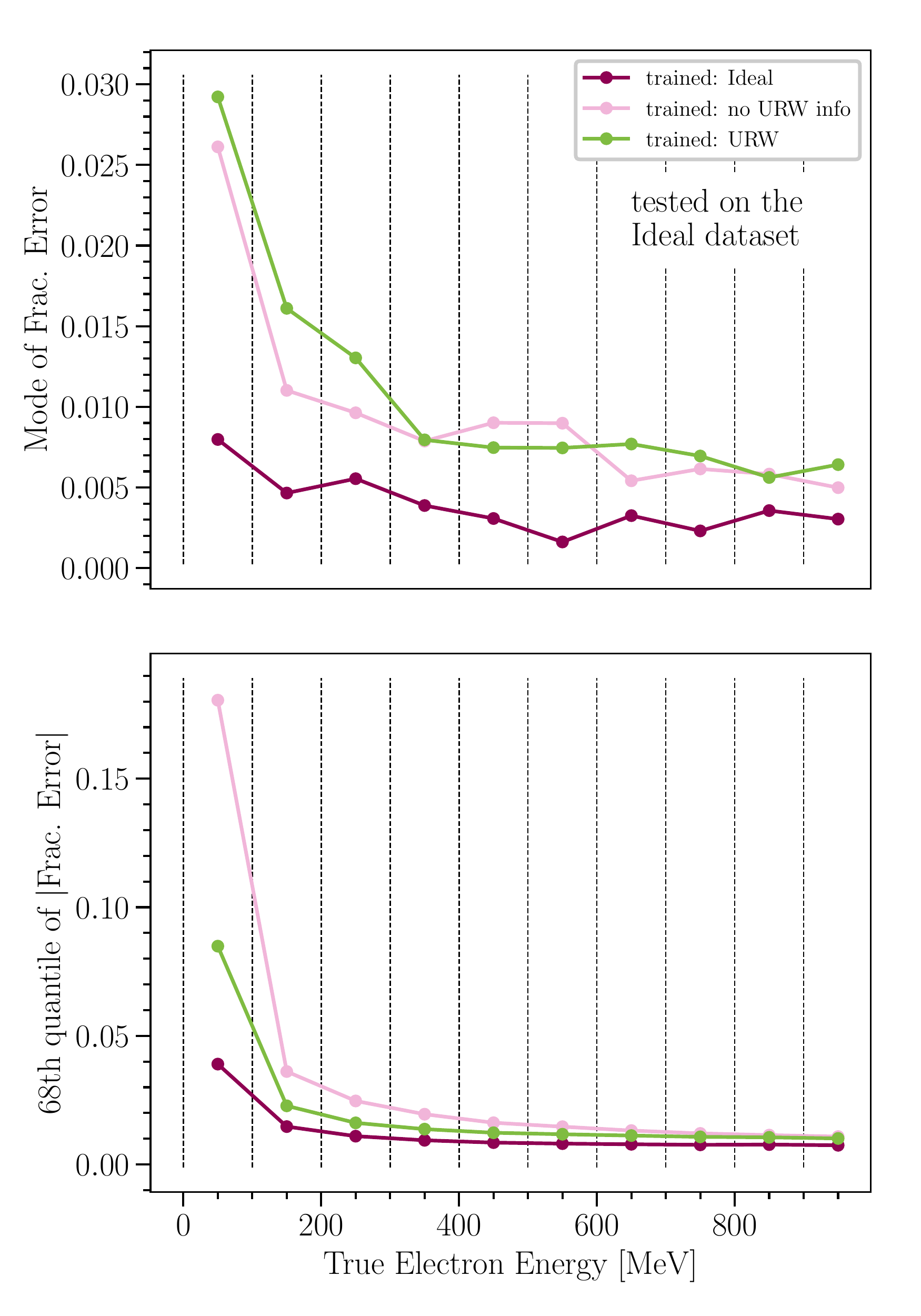}}
\caption{The mode (\textit{above}) and absolute $\SI{68}\percent$ quantiles (\textit{below}) of the fractional reconstruction error, in each bin of true shower energy, for the InceptNet models \textit{(left)} and the ResNet models \textit{(right)}.
These models were trained on three different input datasets, and then tested on ideal events.}
\label{fig:inputs_mpvq_ideal}
\end{figure}

Figure~\ref{fig:inputs_mpvq_ideal} shows the mode and 68\% quantile of each network on the ideal dataset as a function of true energy. The models trained on the unresponsive wire dataset are trained to compensate for missing charge, so on an ideal dataset they overpredict slightly across all shower energies.
We expected this effect to be less significant in the model given wire location information as input, since this information could regulate the compensation for missing charge.
For the ResNet models, this appears to be true: the 68\% quantiles of the model trained without wire information are narrower than those of the model trained with wire information across all shower energies.
This is not clear for the InceptNet models: the model trained without performs similarly, or slightly better than, the InceptNet model trained with information. \\

\begin{figure}[h]
\subfloat{\includegraphics[width=0.5\textwidth]{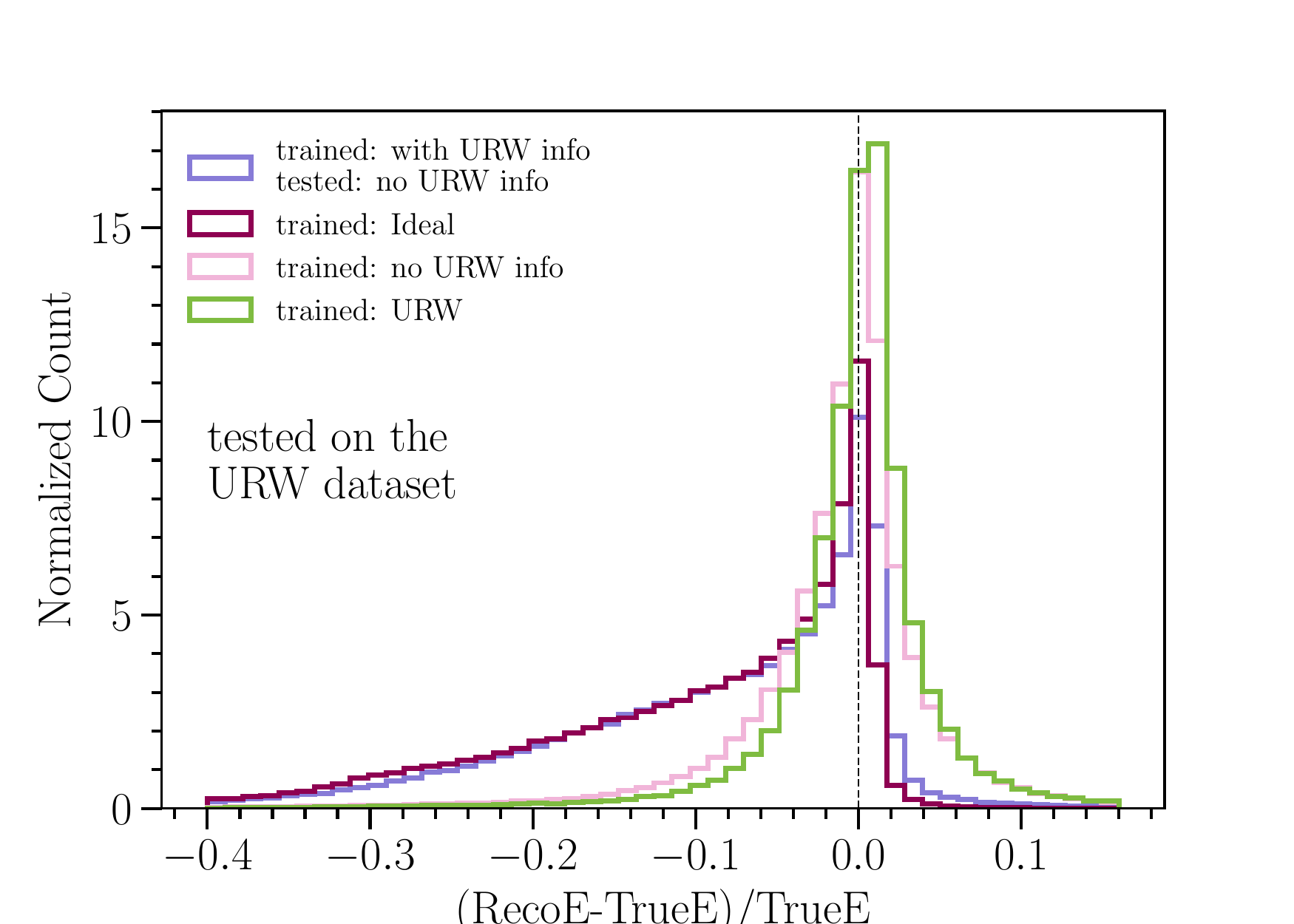}}
\subfloat{\includegraphics[width=0.5\textwidth]{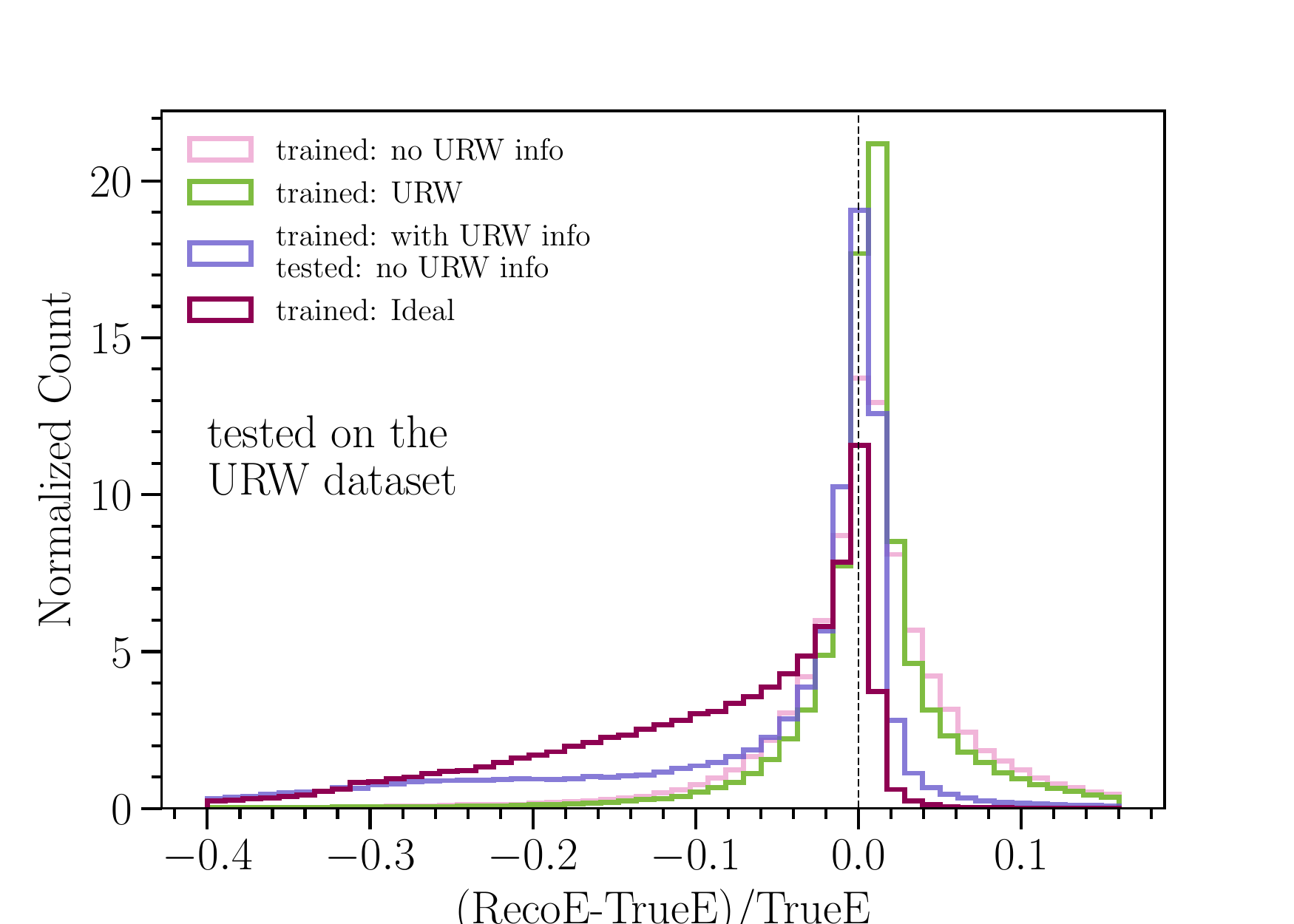}}
\caption{The fractional error of the InceptNet models (\textit{left}) and ResNet models (\textit{right}) trained on three different input datasets, and evaluated on the validation dataset with unresponsive wires (URW).
The performance of the model trained with information on the URW locations, but tested without this information, is plotted in purple-blue.}
\label{fig:inputs_URW}
\end{figure}

Having considered the ideal case, we now turn to the performance of these networks on shower events with unresponsive wires. Figure~\ref{fig:inputs_URW}, like Figure~\ref{fig:inputs_ideal}, shows the fractional error distribution for the InceptNet and ResNet models of the three groups described above, considering now the URW dataset instead of the ideal dataset.
Additionally, we include a fourth class of models which were trained with URW information, but tested without that information (i.e., the wire status input layers are effectively null, incorrectly suggesting to the network that no dead wires exist in the image).
The Ideal models do not reconstruct this dataset well; they underpredict the true energy by more than $\SI{20}\percent$ in over $\SI{18}\percent$ of events.
The models trained with URW information but tested without also do not reconstruct well, and underpredict the true energy by more than $\SI{20}\percent$ in $\SI{14.9}\percent$ and $\SI{16.6}\percent$ of events (for InceptNet and ResNet, correspondingly).
This confirms that these models are taking advantage of the URW information given to them. \\

\begin{figure}[h]
\subfloat{\includegraphics[width=0.5\textwidth]{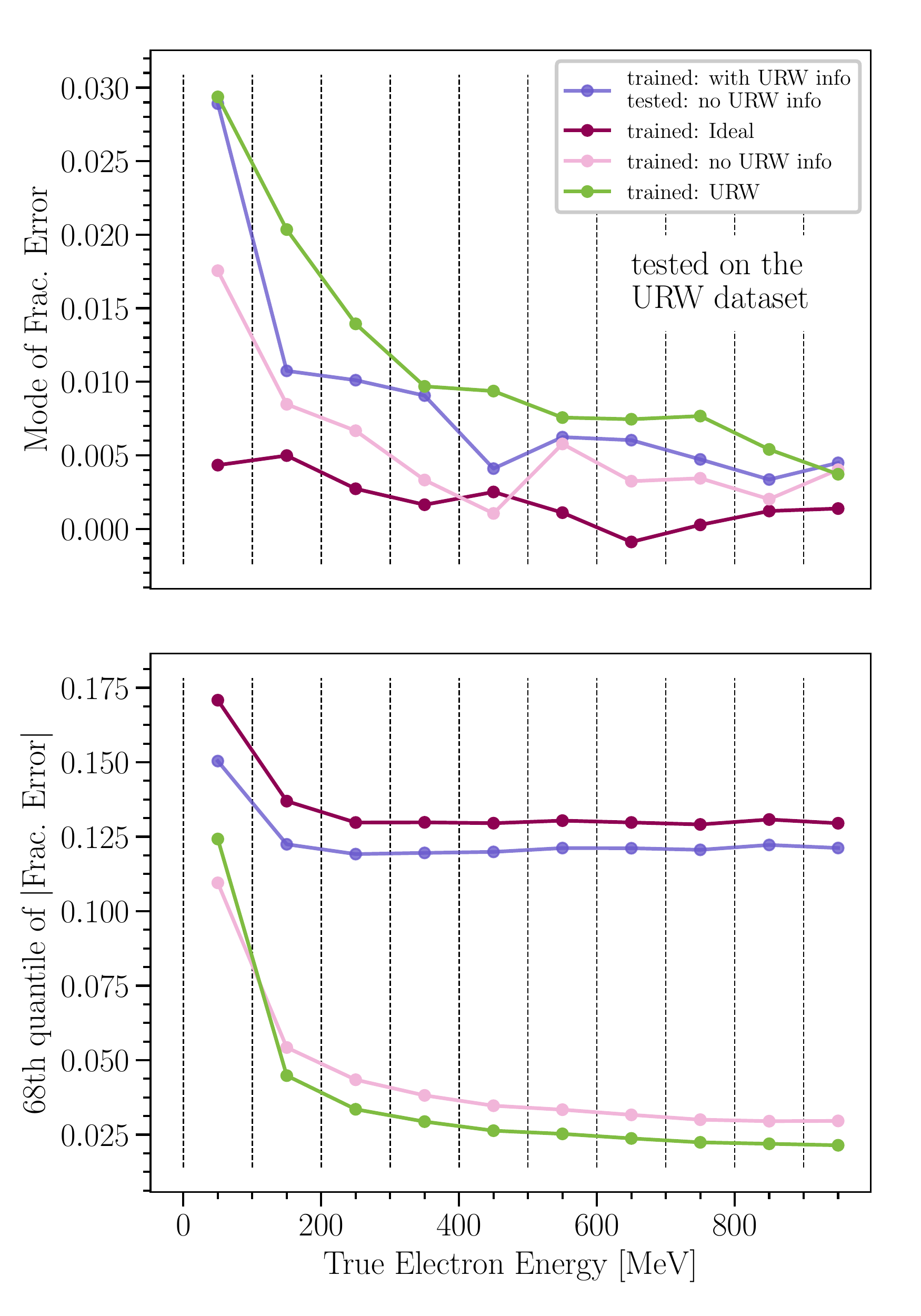}}
\subfloat{\includegraphics[width=0.5\textwidth]{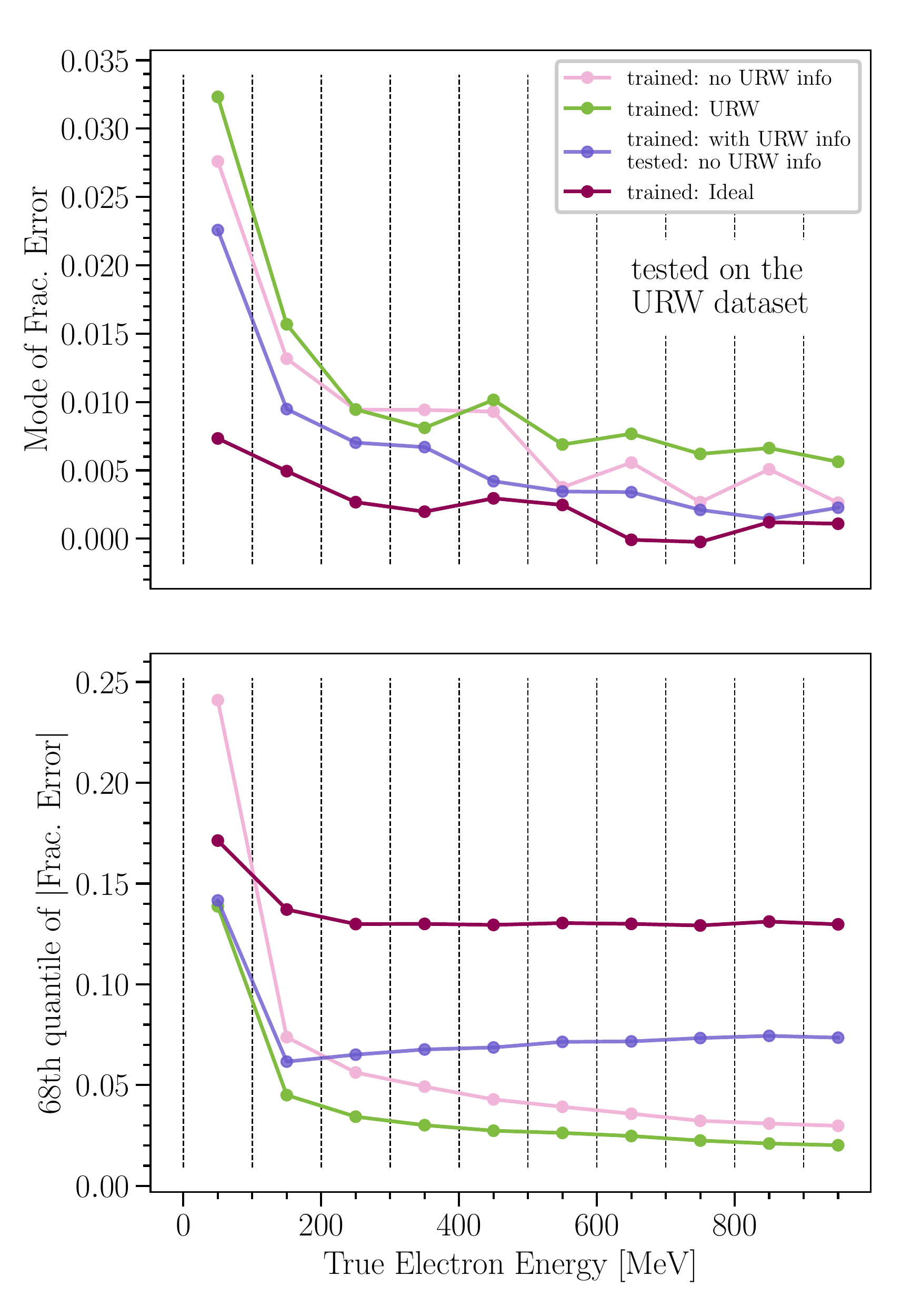}}
\caption{The mode (\textit{above}) and absolute $\SI{68}\percent$ quantiles (\textit{below}) of the fractional reconstruction error, in each bucket of true shower energy, for the InceptNet models \textit{(left)} and the ResNet models \textit{(right)}.
These models were trained on three different input datasets, and then tested on URW events.}
\label{fig:inputs_mpvq_URW}
\end{figure}

Figure~\ref{fig:inputs_mpvq_URW}, like Figure~\ref{fig:inputs_mpvq_ideal}, shows the mode and 68\% quantile of each network as a function of true energy but again considering the URW dataset instead of the ideal datset. Both the InceptNet and ResNet models trained and tested with URW information reconstruct shower energies with much better accuracy than the other models across all shower energies.
They reconstructed $\SI{68}\percent$ of the test events to within $\SI{3.1}\percent$ and $\SI{3.4}\percent$, respectively, and reconstructed $\SI{95}\percent$ of events to within $\SI{14.0}\percent$ and $\SI{16.0}\percent$ accuracy, respectively. This seems to come at the price of a slight tendency to over-estimate shower energies, especially for low energy showers, as shown by the modes in Figure~\ref{fig:inputs_mpvq_URW}. However, the relative increase in the reconstruction accuracy compared to networks trained without URW information significantly outweighs the apparent over-prediction bias.

\subsection{Comparison to the Linear Algorithm's Performance}
\label{sec:linear}

Sections~\ref{section:size},~\ref{section:loss},~and~\ref{section:inputs} narrowed down the best-performing CNN models.
In general, the ``Large'' networks trained under an L1 loss function and provided unresponsive wire information appear to perform optimally, especially on imperfect datasets. 
We now compare these candidates, implemented in both the InceptNet and ResNet architectures, directly to the linear algorithm.
In addition to this comparison between the linear algorithm and neural networks in general, the tests performed in this section are further used to distinguish the performance of the InceptNet and ResNet individually.\\

\begin{figure}[h]
\subfloat{\includegraphics[width=.67\textwidth]{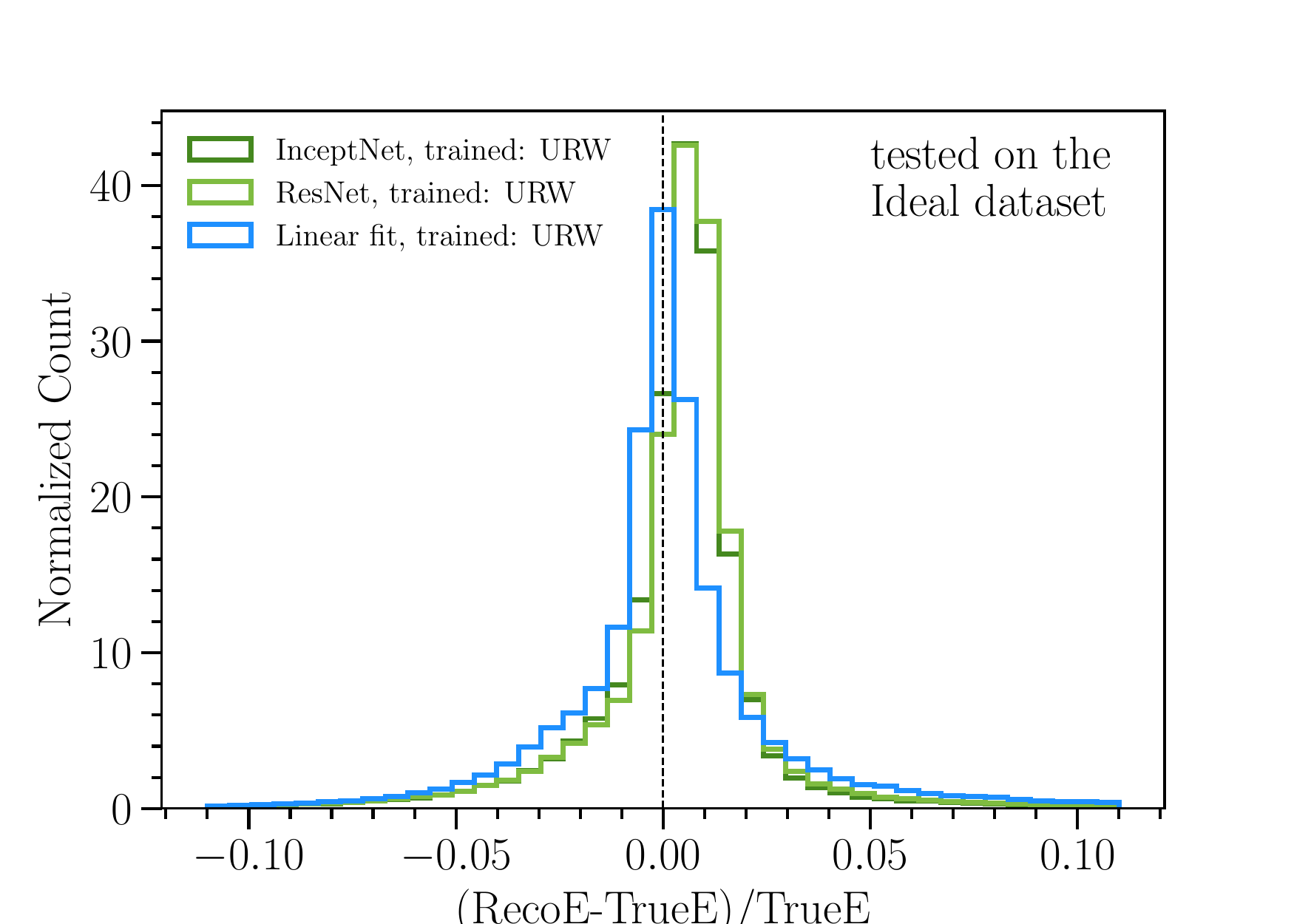}}
\subfloat{\includegraphics[width=.33\textwidth]{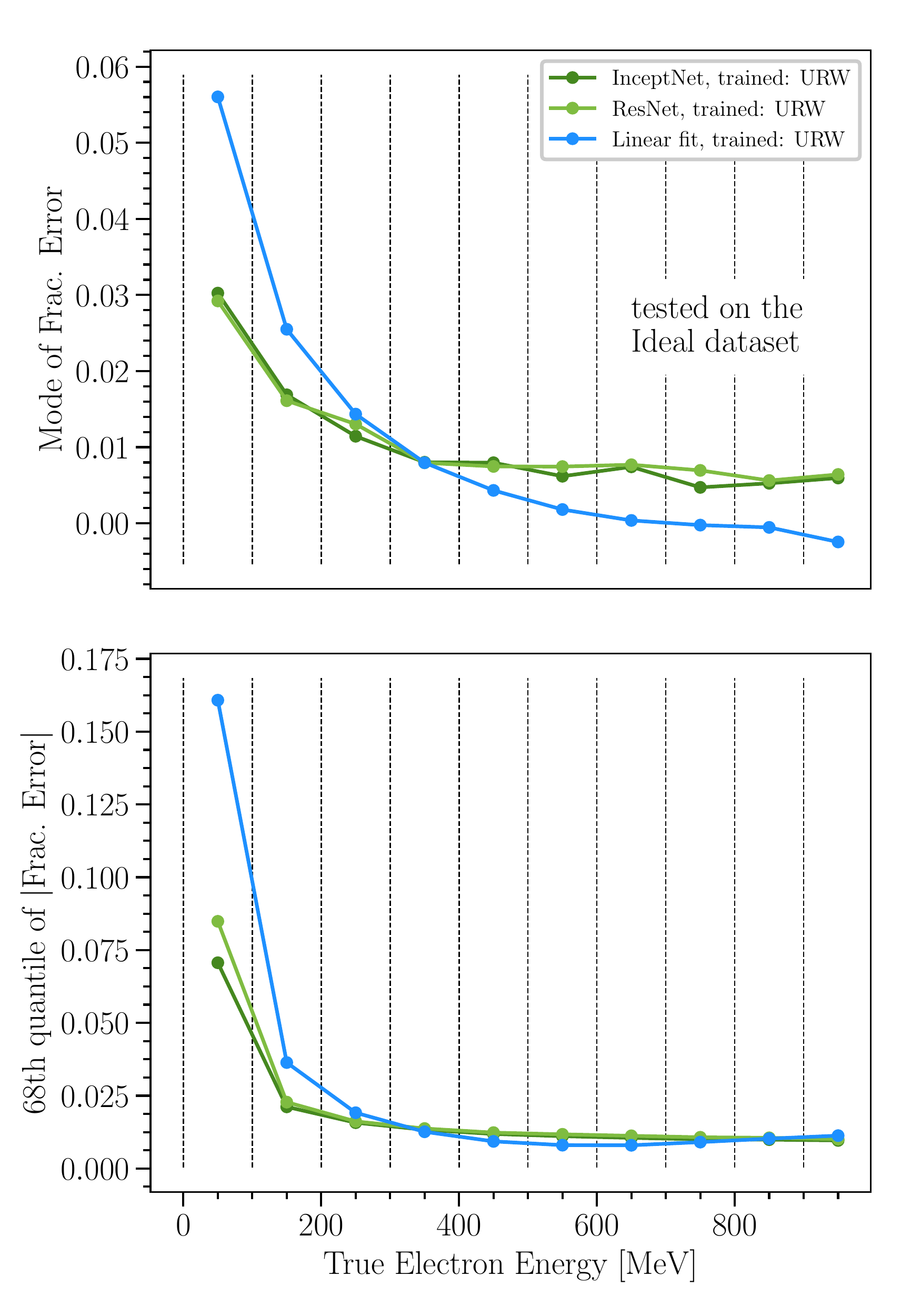}}
\caption{\textit{Left:} The fractional reconstruction error histograms of the linear algorithm and the two CNNs, tested on the ideal validation dataset.
The CNNs were Large sized and trained using the L1 loss function on data with unresponsive wires.
The correction for missing charge due to unresponsive wires may be responsible for the CNN's overprediction bias on these ideal showers.
\textit{Right:} The mode (\textit{above}) and absolute $\SI{68}\percent$ quantiles (\textit{below}) of the fractional reconstruction error for these same two neural networks and the linear algorithm.}
\label{fig:cluster_ideal_traindead}
\end{figure}

In Figure~\ref{fig:cluster_ideal_traindead}, we compare the performance on ideal data of the InceptNet and ResNet models trained on URW data and the linear algorithm fit to the same URW data. 
The linear algorithm, even with suboptimal fit parameters, still produces an excellent reconstruction: the reconstructed energy is overpredicted in $\SI{52.7}\percent$ of events, and $\SI{68}\percent$ of events have a reconstruction error within $\SI{1.99}\percent$.
The neural networks, trained on data with unresponsive wires, slightly over-estimate the energies of these ideal showers: the reconstructed energy is overpredicted by the InceptNet and ResNet models in $\SI{68.9}\percent$ and $\SI{71.2}\percent$ of events respectively.
However, because the neural network fractional error distributions are narrowly peaked, $\SI{68}\percent$ of events are reconstructed to within $\SI{1.43}\percent$ and $\SI{1.52}\percent$, respectively. \\

In most energy bins, the neural networks are able to match the linear algorithm's performance, producing similarly sized confidence intervals and slightly larger modes. In the lowest energy bins, the neural networks actually produce more accurate results, with a mode closer to zero and a smaller confidence interval. While the linear algorithm's calibration procedure sacrifices the fit in the lowest energy bins to secure a good fit at the rest, the neural networks' more flexible reconstruction produces a better compromise overall.  \\

\begin{figure}[h]
\subfloat{\includegraphics[width=.67\textwidth]{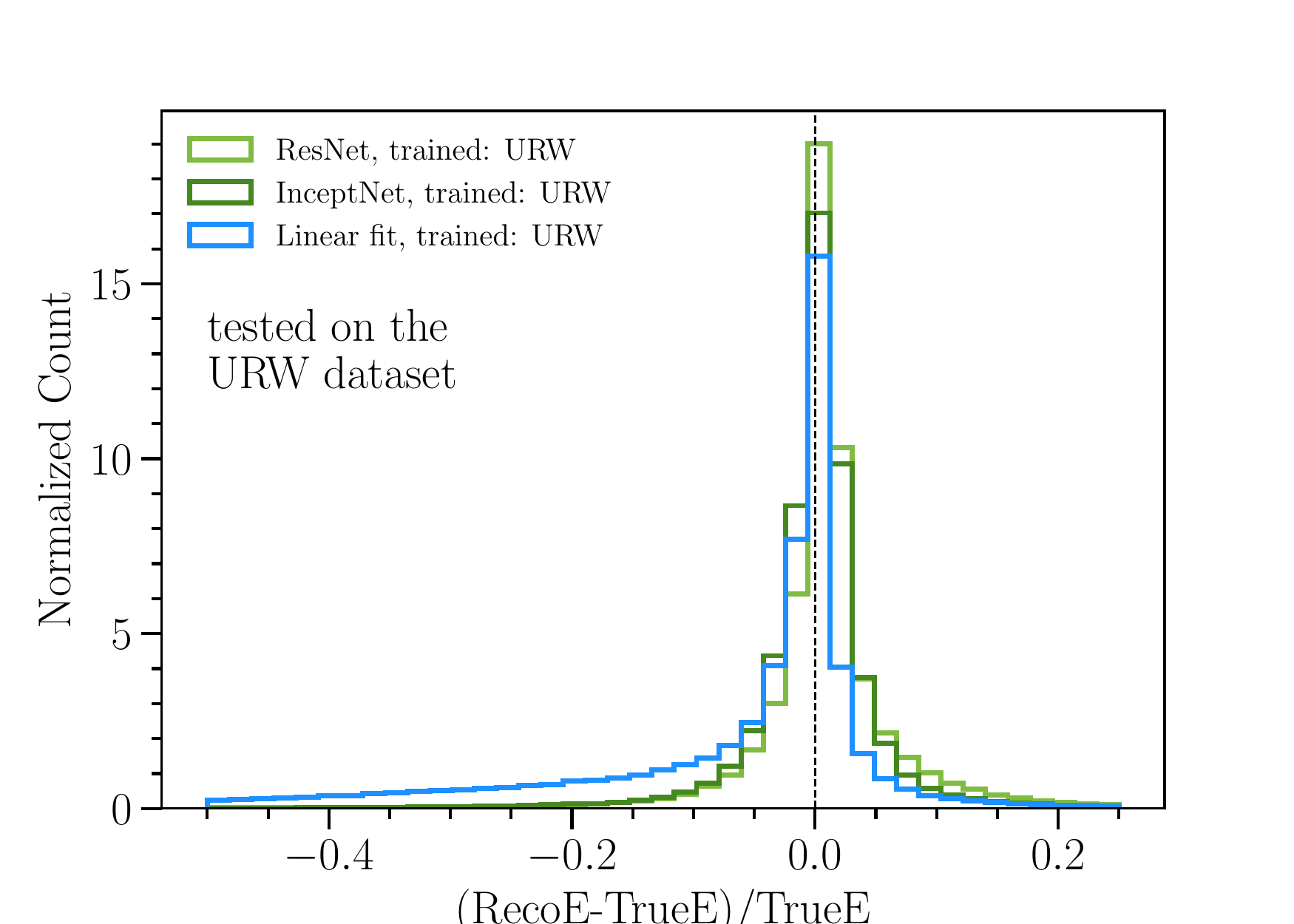}}
\subfloat{\includegraphics[width=.33\textwidth]{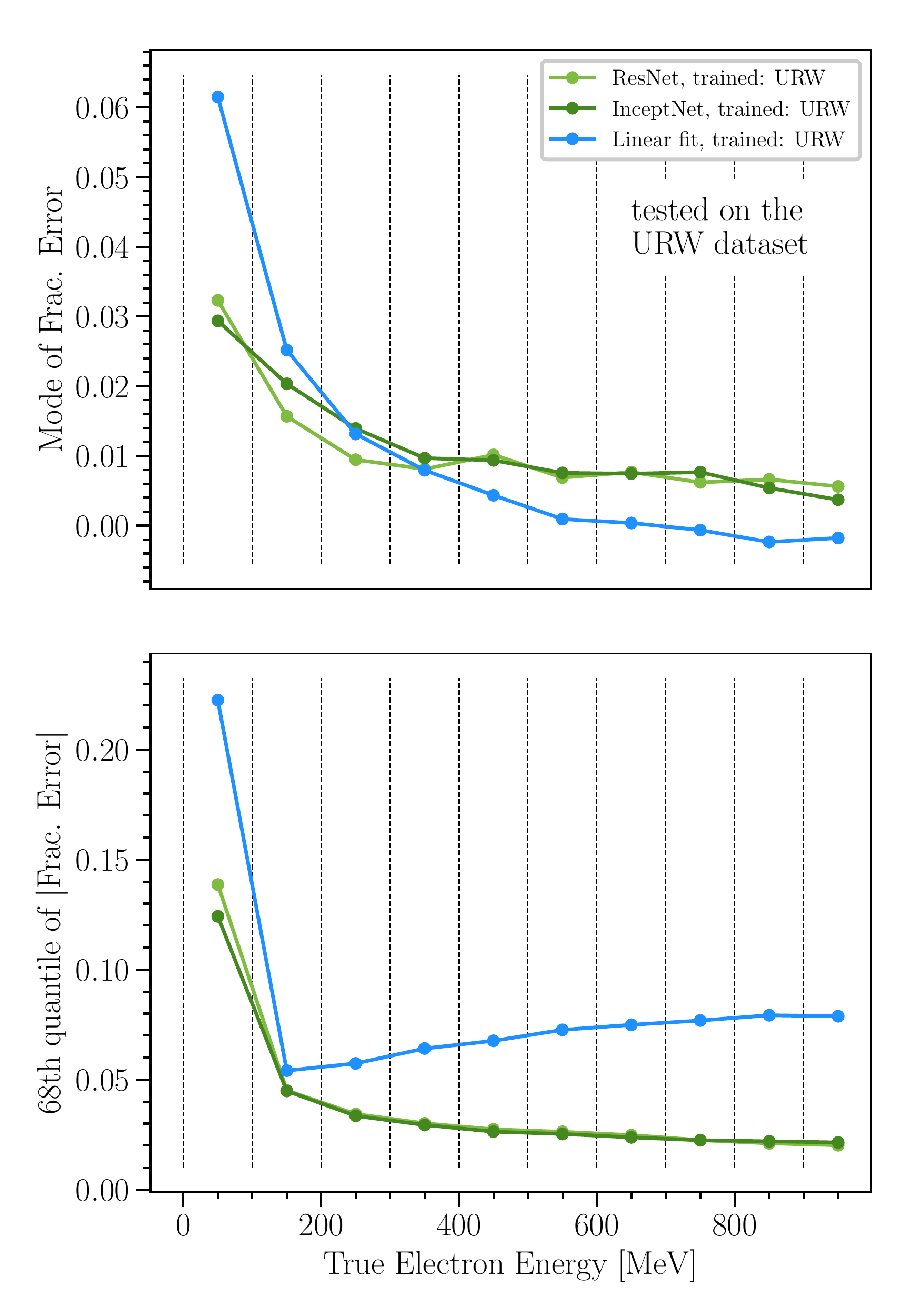}}
\caption{\textit{Left:} The fractional reconstruction error histograms of the linear algorithm and the two CNNs, tested on the validation dataset with unresponsive wires.
The CNNs were Large sized and trained using the L1 loss function on data with unresponsive wires.
\textit{Right:} The mode (\textit{above}) and absolute $\SI{68}\percent$ quantiles (\textit{below}) of the fractional reconstruction error for these same two neural networks and the linear algorithm.
All three plots demonstrate the CNNs' robustness to unresponsive wire effects. }
\label{fig:cluster_URW}
\end{figure}

For the more realistic dataset with unresponsive wires, the neural networks and the linear algorithm offer different advantages.
The performance of each on this dataset is compared in Figure \ref{fig:cluster_URW}. 
The fractional error distributions for the InceptNet and ResNet models are still slightly asymmetric: $\SI{66.4}\percent$ and $\SI{56.9}\percent$ of events are overpredicted, respectively.
The fractional error distribution for the linear algorithm has a long tail of underpredictions, over-estimating the energy of only $\SI{32.4}\percent$ of events (i.e. under-estimating the energy of $\SI{67.6}\percent$). 
The linear algorithm can provide very accurate reconstructions for a larger fraction of events: it reconstructs $\SI{18.4}\percent$ of the total dataset to within $\SI{0.5}\percent$ error, while InceptNet and ResNet models manage $\SI{15.9}\percent$ and $\SI{15.5}\percent$ of the total dataset, respectively.
By the $\SI{1}\percent$ error cutoff, the neural networks just barely overpass the linear algorithm: the linear algorithm reconstructs 29.6\%, the InceptNet model 30.8\%, and the ResNet model 32.2\% of the total dataset to within $\SI{1}\percent$ accuracy.\\

When considering reconstruction on the bulk of events, however, the neural networks are the better choice.
The InceptNet and ResNet models reconstructed $\SI{68}\percent$ of the test dataset to within $\SI{3.40}\percent$ and $\SI{3.15}\percent$ accuracy, respectively.
The linear algorithm's interval is more than double that: $\SI{68}\percent$ of events are contained within $\SI{8.9}\percent$ fractional error.
For $\SI{95}\percent$ of events, the fractional errors of the InceptNet and ResNet reconstructions are within $\SI{14.91}\percent$ and $\SI{18.49}\percent$, respectively, while the same bound for the linear algorithm reconstruction is $\SI{53}\percent$.
When the linear algorithm fails to reconstruct an event, presumably because it passed through a dense region of unresponsive wires, it fails thoroughly. \\

\begin{wrapfigure}{L}{0.5\textwidth}
\includegraphics[width=\textwidth]{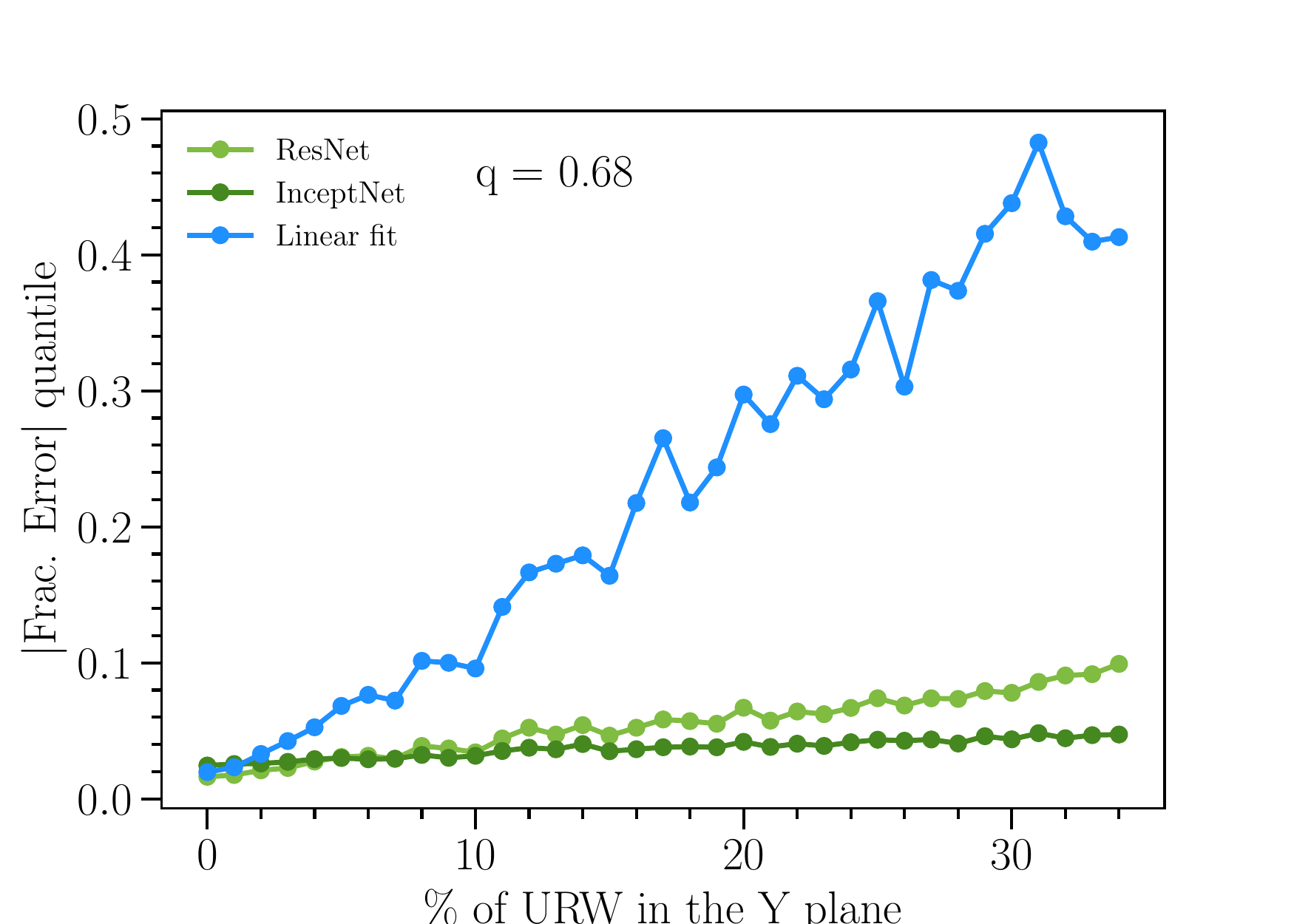}
\caption{The absolute $\SI{68}\percent$ quantile of the fractional error, plotted as a function of the percentage of unresponsive wires affecting the Y plane of the input images.}
\label{fig:cluster_URW_q}
\end{wrapfigure}

\begin{figure}[h!]
\subfloat{\includegraphics[width=.49\textwidth]{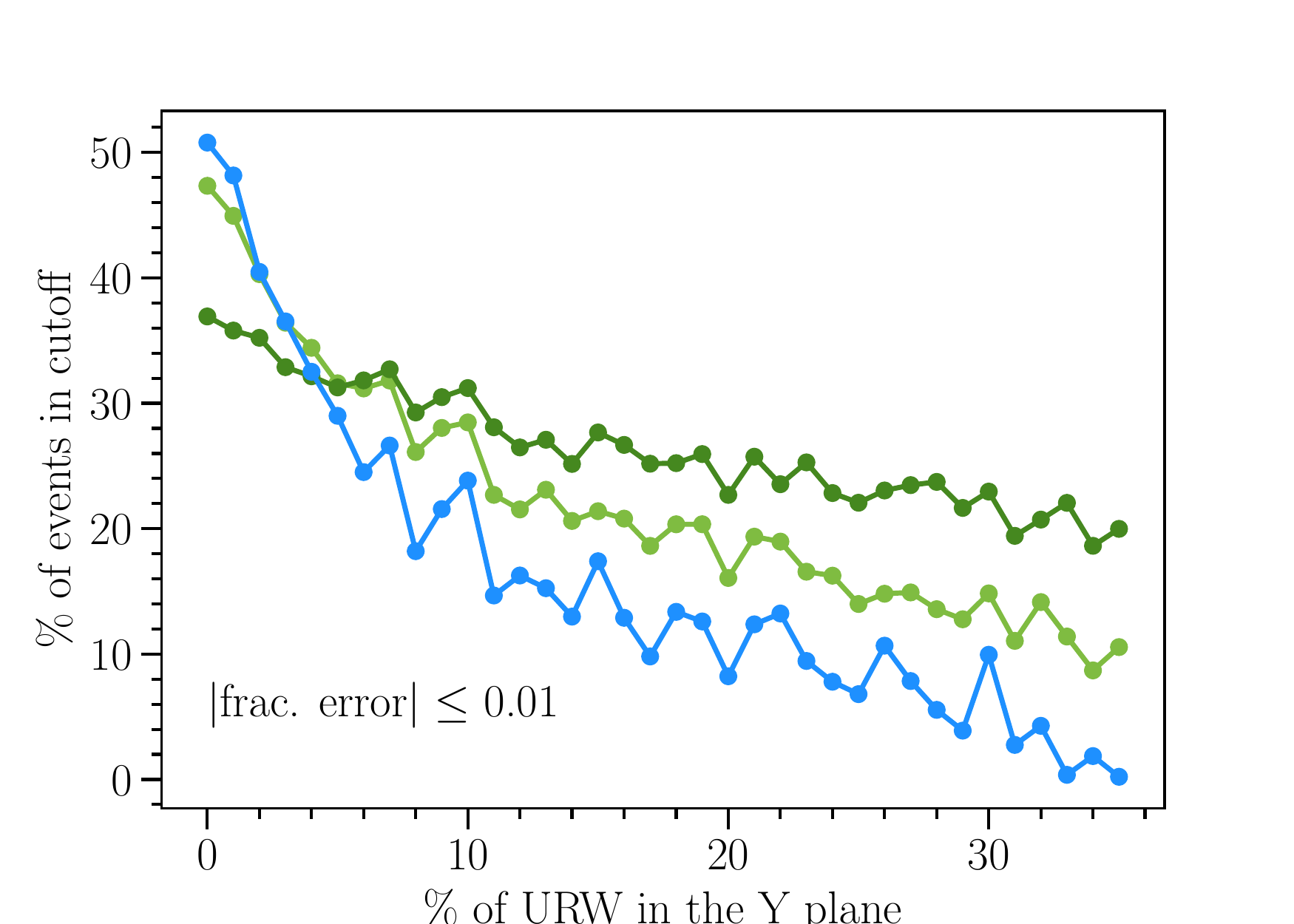}}
\subfloat{\includegraphics[width=.49\textwidth]{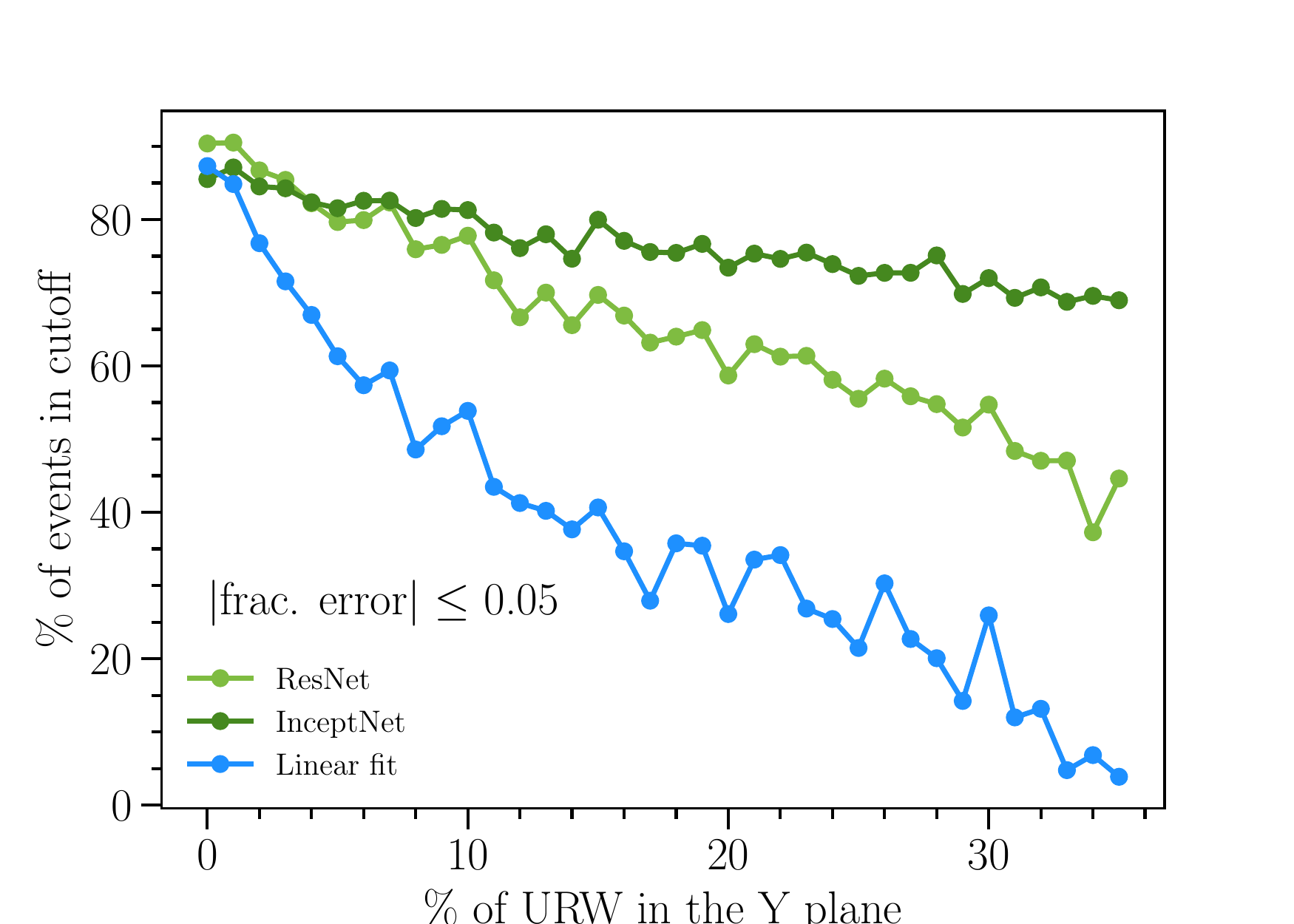}}
\caption{The percentage of events reconstructed to within $\SI{1}\percent$ (\textit{left}) or $\SI{5}\percent$ (\textit{right}) accuracy, as a function of the percentage of unresponsive wires affecting the Y plane of the input images.
The linear algorithm reconstructs a larger fraction of almost-ideal events within the stricter $\SI{1}\percent$ cutoff than the CNNs, but a smaller fraction of these events within the wider $\SI{5}\percent$ cutoff.}
\label{fig:cluster_URW_c}
\end{figure}

For one-third of our test dataset, the electron showers don't pass through any unresponsive wires in the Y-plane.
On these events, the linear algorithm is extremely accurate.
On the remaining two-thirds of events, however, its performance declines much more significantly than that of the neural networks.
Figures~\ref{fig:cluster_URW_q} and~\ref{fig:cluster_URW_c} show this effect from two different points of view.
In Figure~\ref{fig:cluster_URW_q} we show the $\SI{68}\percent$ quantile of the absolute fractional error as a function of the percentage of unresponsive wires in an event.
Events with more than $\SI{10}\percent$ unresponsive wires are very poorly reconstructed by the linear algorithm, and contribute heavily to the long tail of the fractional error histogram.
From Figure~\ref{fig:cluster_URW_q}, one can see that events with at least $\SI{10}\percent$ unresponsive wires have energy resolution $>\SI{10}\percent$ in the linear algorithm, while the energy resolution of the neural network models is considerable better in this regime.
One can also see that InceptNet appears to be more robust to larger URW fractions than ResNet. 
In Figure~\ref{fig:cluster_URW_c} we plot the percentage of events reconstructed to within $\SI{1}\percent$ and $\SI{5}\percent$ accuracy.
Interestingly, for a URW fraction $\lesssim \SI{5}\percent$, the linear algorithm and the ResNet algorithm reconstruct events to $\SI{1}\percent$ accuracy than the InceptNet algorithm.
This trend reverses for events with $>\SI{5}\percent$ unresponsive wires: the InceptNet algorithm reconstructs more of these events to $\SI{1}\percent$ accuracy than the ResNet and linear algorithms, though the ResNet algorithm still outperforms the linear algorithm in this regime.
When the error interval expands from $\SI{1}\percent$ to $\SI{5}\percent$, both neural network architectures outperform the linear algorithm, even on events with fewer than $\SI{5}\percent$ unresponsive wires in the Y-plane.
As the URW fraction increases, the neural networks are able to reconstruct a significantly higher fraction of events to within $\SI{5}\percent$ accuracy compared with the linear algorithm.
Thus, one can conclude that shower energy reconstruction algorithms which use CNNs are considerably more robust to showers passing with a large fraction of unresponsive wires compared to traditional linear calibration algorithms. \\

Additionally, the same trend is observed in the $\SI{5}\percent$ accuracy case when comparing the two network architectures: the InceptNet algorithm outperforms the ResNet algorithm for large URW fractions, while the ResNet algorithm performs better in the very small URW fraction regime ($\lesssim \SI{5}\percent$ of wires).
As the ResNet algorithm behaves more similarly than the InceptNet algorithm to the linear algorithm, it may be striking a balance between peak precision and bulk accuracy. \\

Therefore, the optimal network architecture for reconstructing shower energies in a LArTPC depends in general on the goal of the analysis and more specifically on the anticipated fraction of unresponsive wires through which the showers travel.

\section{Conclusion}
\label{section:conclusion}

In this report, we studied the ability of CNN-based algorithms to reconstruct electromagnetic shower energies in a LArTPC.  
Different classes of CNN algorithms were compared in their ability to reconstruct showers accurately and with minimal bias across a range of energies.
These studies indicate that CNN-based shower energy reconstruction algorithms in a LArTPC are robust against showers that pass through multiple chunks of unresponsive wires in the detector.
Their competitiveness with typical clustering linear algorithms owes to the ability to reconstruct a larger fraction of events with moderate amounts of missing charge to within reasonable ($\lesssim \SI{5}\percent$) accuracy.
We found that CNNs could and did utilize information on the location of unresponsive wires to further improve reconstruction efficiency.
Additionally, we found that the InceptNet architecture is slightly more robust than the ResNet architecture to showers that pass through larger fractions of unresponsive wires, though the latter performs better than the former in cases of few unresponsive wires.\\

The performance of the clustering linear algorithm in these results was in one sense optimal: since our simulated events had no background from non-shower charge depositions, there was no inefficiency due to charge being miscategorised as shower or not.
We therefore expect that on more realistic datasets, the relative performance of the CNNs could further improve. \\

Larger CNNs, with more parameters, reconstructed the majority of events to within smaller error.
In addition, the loss curves over the training epochs of all our CNNs indicate that they could still achieve slight further improvements with additional training time.
Thus, the optimal performance of the CNNs is constrained by cost considerations in memory and time.
Still, the main takeaway from this study is clear: CNN-based shower energy reconstruction algorithms in LArTPCs show significant improvement over traditional linear reconstruction algorithms on showers that pass through unresponsive regions of the detector.

\acknowledgments
NSF grant PHY-1801996 supported 
KC, JMC, AS and NWK for this work. Additionally, KC is supported by the Mariana Polonsky Slocum (1955) Memorial Fund. This material is based upon work supported by the National Science Foundation Graduate Research Fellowship under Grant No. 1745302. We thank Kazuhiro Terao and John Hardin for useful discussions.

\pagebreak{}
\appendix

\section{Architecture Details}
\label{app:archs}
\begin{figure}[h]
    \caption{Larger Residual Networks were designed by composing more layers out of more basic residual blocks.}
    \subfloat{\includegraphics[width=\textwidth]{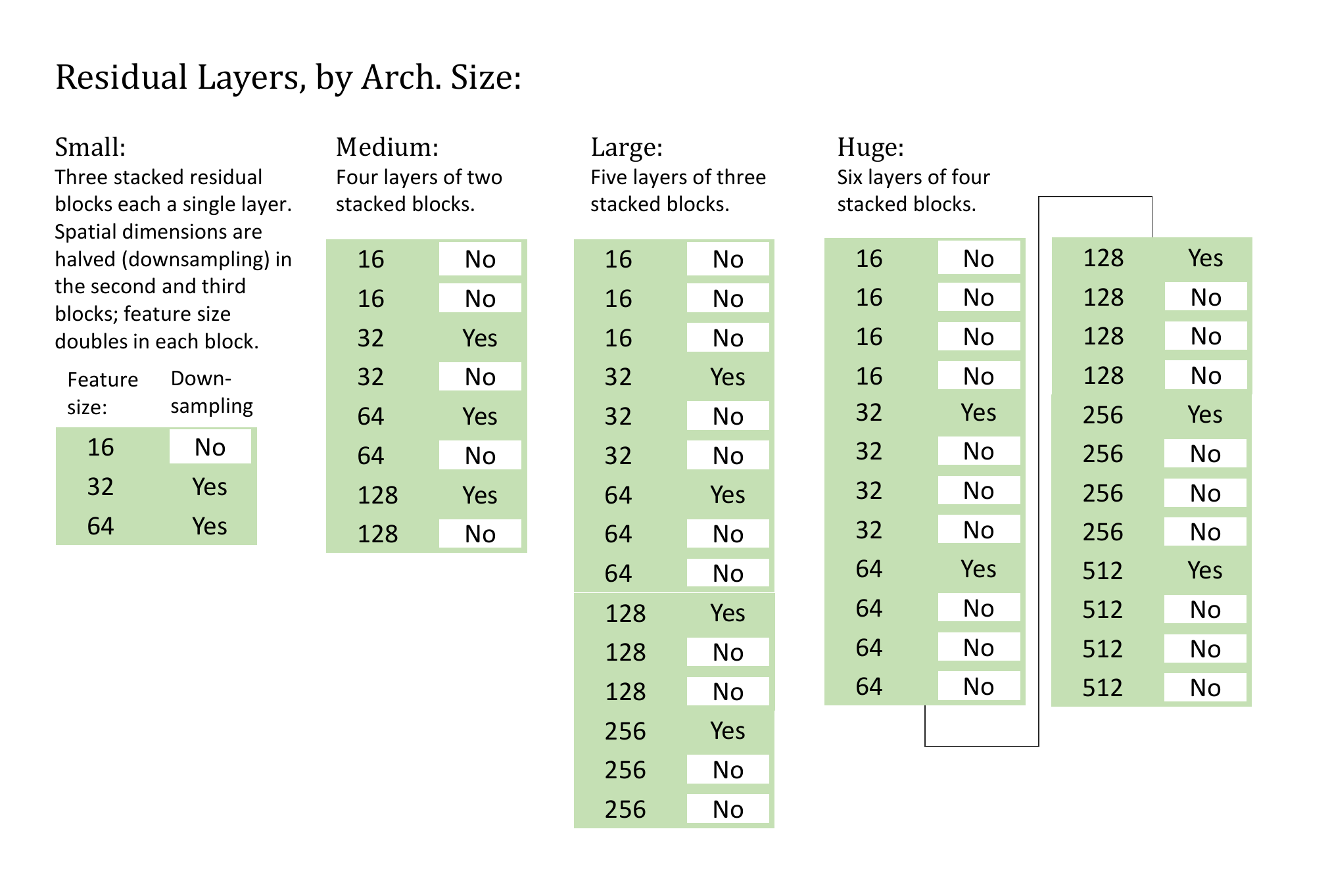}}
    \label{fig:arch_res_sizes}
\end{figure}

\begin{figure}[h]
    \caption{Larger Inception Networks were designed by adding more basic inception blocks to each layer, and by increasing the output feature sizes of all convolutional filters throughout.}
    \subfloat{\includegraphics[width=\textwidth]{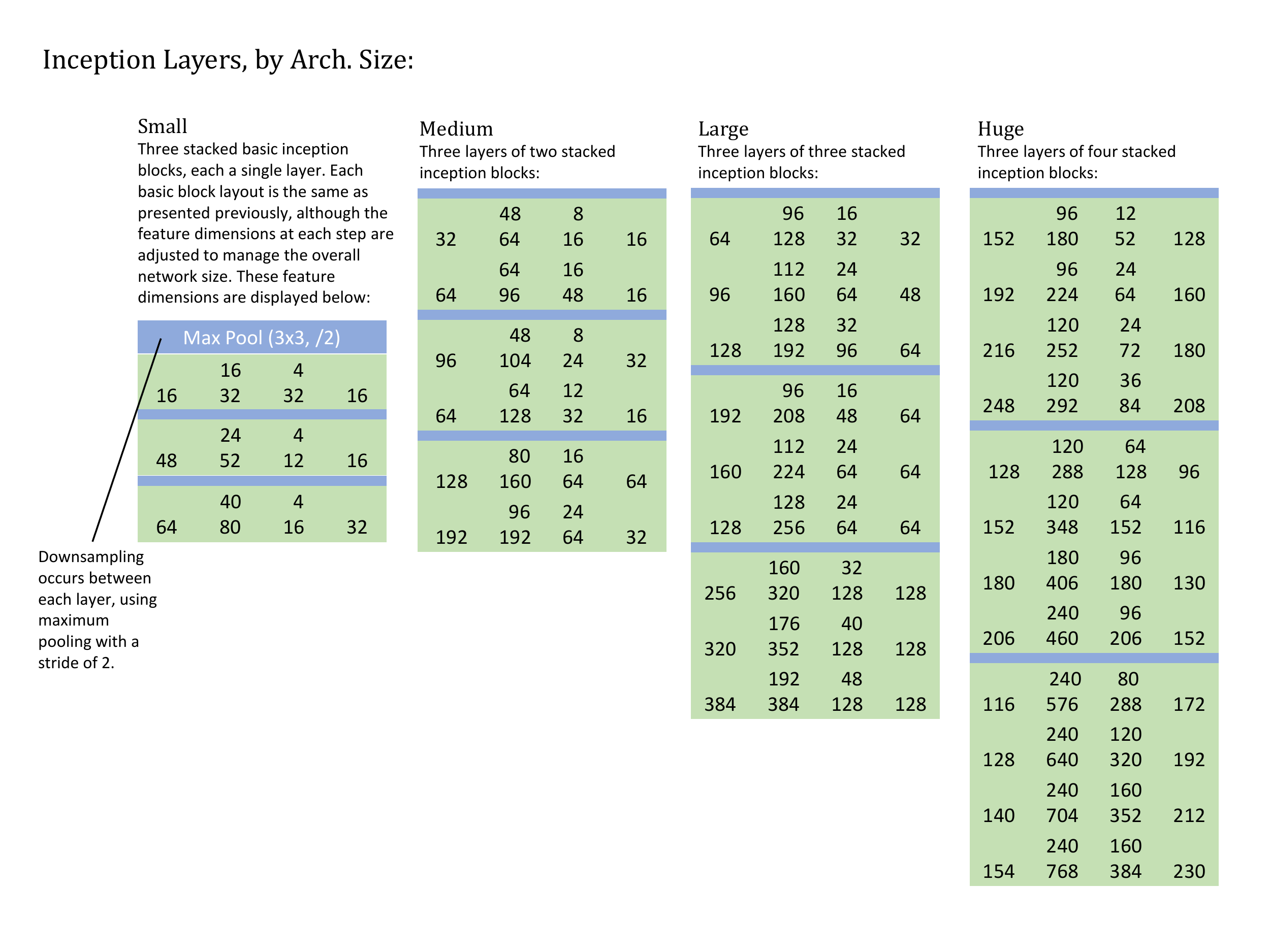}}
    \label{fig:arch_incept_sizes}
\end{figure}

Figures \ref{fig:arch_res_sizes} and \ref{fig:arch_incept_sizes} schematically depict the convolutional layers of the Residual and Inception networks at each scale. 

\section{Further Detail on the Studies}
\label{app:details}

In this section, we provide further quantitative detail on the different studies described throughout this paper.

\begin{figure}[h]
\subfloat{\includegraphics[width=0.5\textwidth]{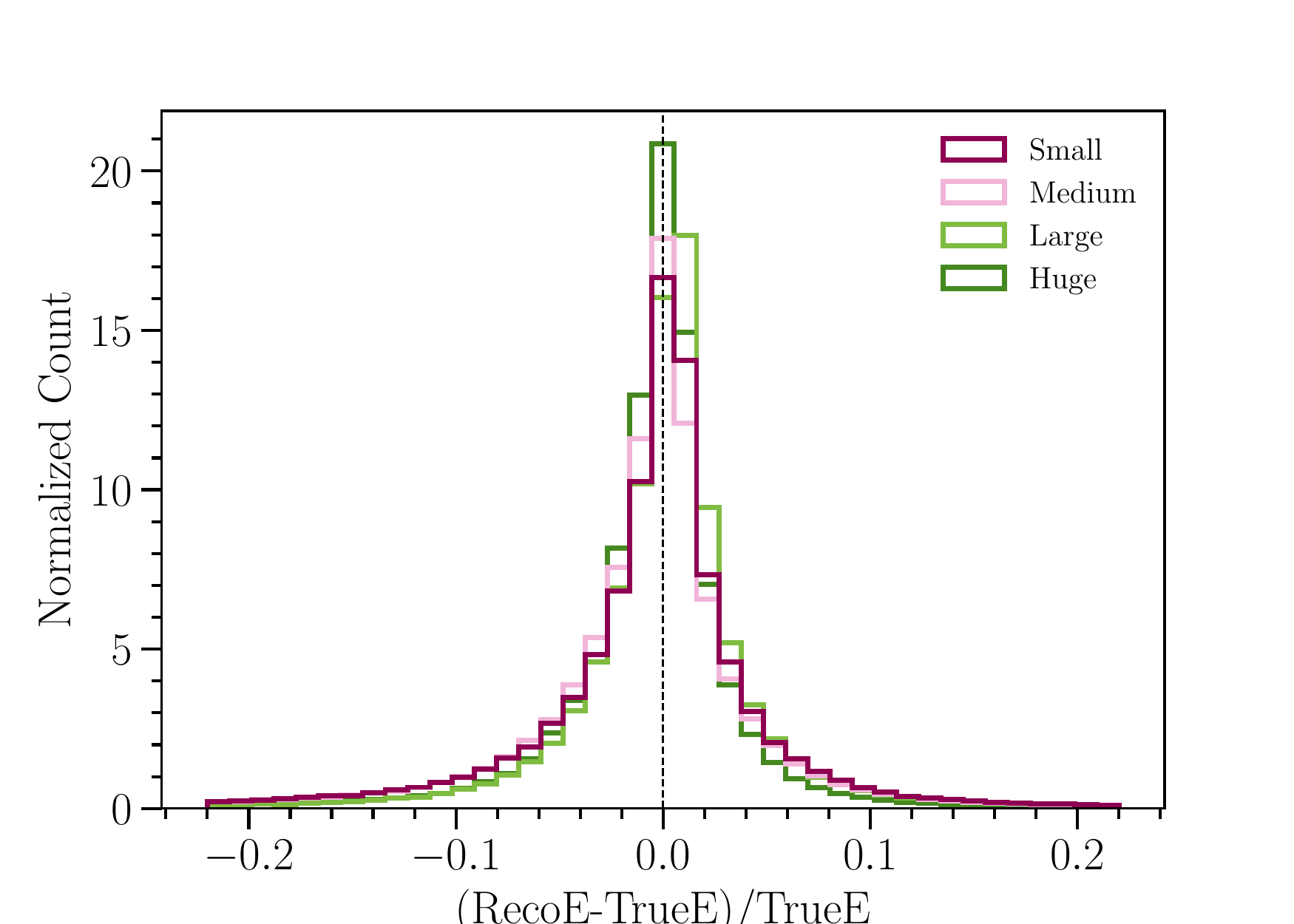}}
\subfloat{\includegraphics[width=0.5\textwidth]{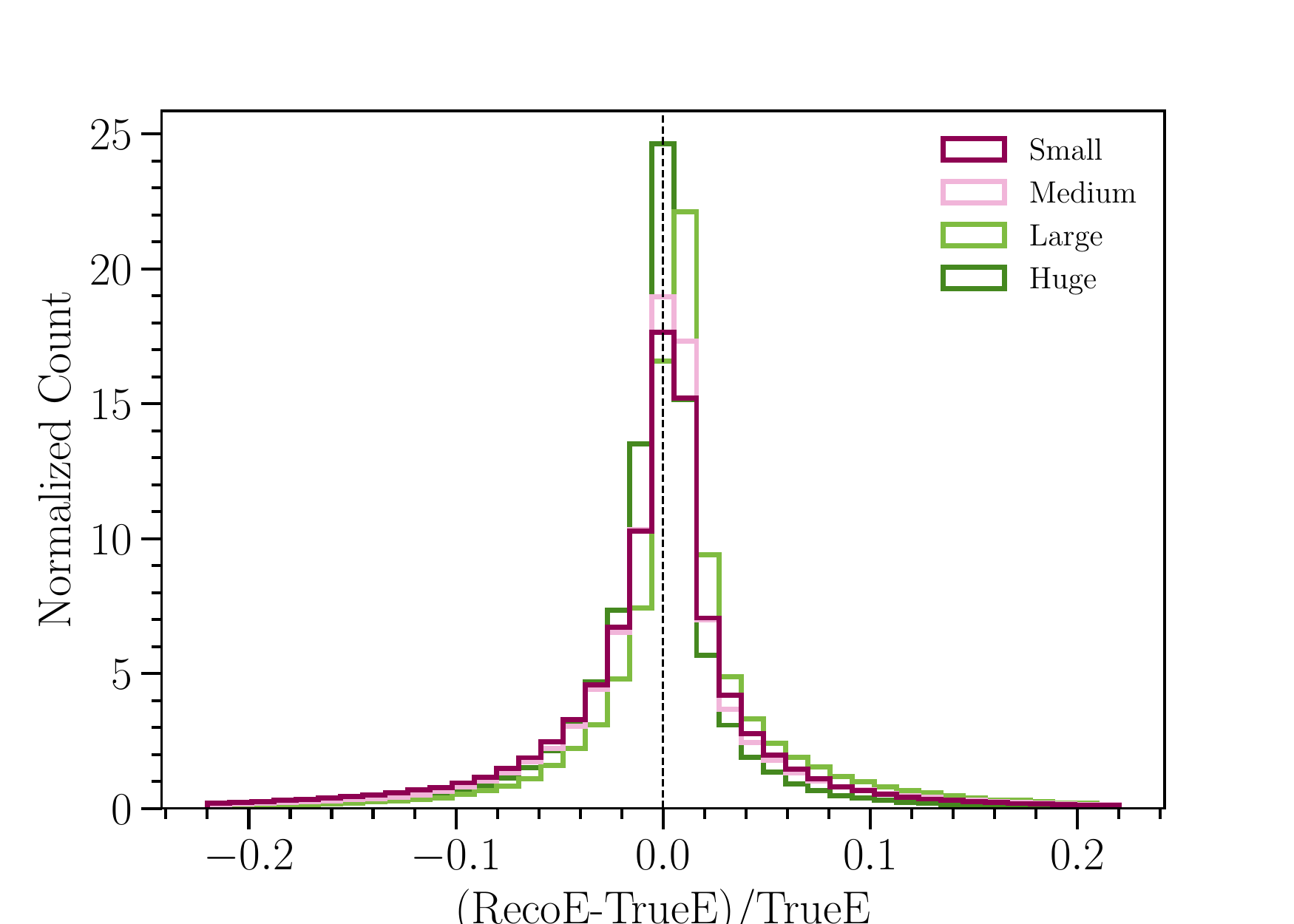}}
\caption{The normalized distributions of the fractional reconstruction error for the different sized InceptNet (\textit{left}) and ResNet (\textit{right}) models.}
\label{fig:size_hist}
\end{figure}

We first discuss the comparison between networks of different sizes, described in Section~\ref{section:size}. The fractional reconstruction error distribution for each size model in Table~\ref{table:model_params} is plotted in Figure~\ref{fig:size_hist}.
The widths of the distributions get narrower as the models increase in size, as expected. Table~\ref{table:size_hist} reports additional quantitative information on these distributions. Specifically, we provide (1) the fraction of events with over-estimated energies (expected to be $\sim 50$\% for an unbiased reconstruction), (2) the most probable fractional error (expected to be $\sim 0$\% for an unbiased estimator), and (3) the 68\% quantile for each fractional error distribution (i.e., the $1\sigma$ uncertainty on the energy estimation).

We next consider the loss function comparison described in Section~\ref{section:loss}. Additional quantitative information concerning the fractional error distributions shown in Figure~\ref{fig:loss_hist} is provided in Table~\ref{table:loss_hist}. This table confirms the conclusion in the main text; the model trained using an L1 loss function performs slightly better than the model trained under an MSE loss function, while the model trained with a fractional error loss function seems to significantly over-predict the energy of most showers. 

In Table~\ref{table:inputs_hist} we provide additional  quantitative information on the fractional error distributions in Figures~\ref{fig:inputs_ideal} and \ref{fig:inputs_URW}. This table again confirms the conclusions in the main text; when estimating energies of showers without URWs, the networks trained on events with URWs tend to over-predict compared to the networks trained on ideal events. However, the networks trained on images with URWs perform much better on events which also include URWs. This is especially true when information on the URWs are provided as input to the networks. It is also clear that the networks trained with input information on URW are making use of this information, as the prediction uncertainty (given by 68\% quantiles in Table~\ref{table:inputs_hist}) increases when this input is removed.

Finally, we report in Table~\ref{table:cluster_hist} additional quantitative information of the fractional error distributions shown in Figure~\ref{fig:cluster_ideal_traindead} and \ref{fig:cluster_URW}. 
This table considers the performance of the linear algorithm and the neural networks on evaluation datasets simulating both ideal and imperfect detectors. As discussed in Section~\ref{sec:linear}, one can see from the table that the neural network algorithms are in general more robust to unresponsive wires introduced in the imperfect dataset.

\begin{table}[h]
\begin{centering}
\begin{tabular}{l l l l l}
\toprule
\makecell[tl]{Architecture} & \makecell[tl]{Size} & \makecell[tl]{\% > 0} & \makecell[tl]{Mode\\(\% error)} & \makecell[tl]{$\SI{68}\percent$ quantile\\(\% error)} \\
\midrule
InceptNet & Small    & 50.2  & 0.52 & 3.96 \\
InceptNet & Medium   & 45.5  & 0.14 & 3.86 \\
InceptNet & Large    & 56.9  & 0.71 & 3.15 \\
InceptNet & Huge     & 47.6  & 0.25 & 2.79 \\
ResNet & Small    & 50.7     & 0.46 & 3.89 \\
ResNet & Medium   & 53.2     & 0.56 & 3.48 \\
ResNet & Large    & 66.4     & 0.78 & 3.40 \\
ResNet & Huge     & 47.1     & 0.32 & 2.56 \\
\bottomrule
\end{tabular}
\caption{Quantitative properties of the fractional error histograms in Figure~\ref{fig:size_hist}. See text for a description of the variables shown in each column.}
\label{table:size_hist}
\end{centering}
\end{table}

\begin{table}[h]
\begin{centering}
\begin{tabular}{l l l l l}
\toprule
\makecell[tl]{Architecture} & \makecell[tl]{Loss} & \makecell[tl]{\% > 0} & \makecell[tl]{Mode\\(\% error)} & \makecell[tl]{$\SI{68}\percent$ quantile\\(\% error)} \\
\midrule
InceptNet & L1     & 56.9  & 0.71 & 3.15 \\
InceptNet & Frac.  & 45.6  & 5.04 & 8.56 \\
InceptNet & MSE    & 59.8  & 0.95 & 3.25 \\
ResNet & L1      & 66.4     & 0.78 & 3.40 \\
ResNet & Frac.   & 45.5     & 6.07 & 8.46 \\
ResNet & MSE     & 61.4     & 0.89 & 4.23 \\
\bottomrule
\end{tabular}
\caption{Quantitative properties of the fractional error histograms in Figure \ref{fig:loss_hist}. }
\label{table:loss_hist}
\end{centering}
\end{table}

\begin{table}[h]
\begin{centering}
\begin{tabular}{l l l l l l}
\toprule
\makecell[tl]{Architecture} & \makecell[tl]{Training Data} & \makecell[tl]{Testing Data} & \makecell[tl]{\% > 0} & \makecell[tl]{Mode\\(\% error)} & \makecell[tl]{$\SI{68}\percent$ quantile\\(\% error)} \\
\midrule
InceptNet & Ideal        & Ideal     & 49.8  & 0.36 & 1.05 \\
InceptNet & URW, no info & Ideal     & 62.3  & 0.55 & 1.36 \\
InceptNet & URW          & Ideal     & 68.9  & 0.81 & 1.43 \\
ResNet    & Ideal        & Ideal     & 49.9  & 0.37 & 1.05 \\
ResNet    & URW, no info & Ideal     & 71.4  & 0.57 & 1.97 \\
ResNet    & URW          & Ideal     & 71.2  & 0.70 & 1.15 \\
InceptNet & Ideal        & URW           & 12.7  & 0.11 & 13.33 \\
InceptNet & URW, no info & URW           & 46.3  & 0.33 & 3.90 \\
InceptNet & URW          & URW           & 56.9  & 0.71 & 3.14 \\
InceptNet & URW          & URW, no info  & 20.3  & 0.41 & 12.33 \\
ResNet    & Ideal        & URW           & 12.7  & 0.11 & 13.33 \\
ResNet    & URW, no info & URW           & 60.5  & 0.35 & 4.87 \\
ResNet    & URW          & URW           & 66.4  & 0.78 & 3.40 \\
ResNet    & URW          & URW, no info  & 33.6  & 0.34 & 7.86 \\
\bottomrule
\end{tabular}
\caption{Quantitative properties of the fractional error histograms in Figure~\ref{fig:inputs_ideal} and Figure~\ref{fig:inputs_URW}}
\label{table:inputs_hist}
\end{centering}
\end{table}

\begin{table}[h]
\begin{centering}
\begin{tabular}{l l l l l l}
\toprule
\makecell[tl]{Architecture} & \makecell[tl]{Training Data} & \makecell[tl]{Testing Data} & \makecell[tl]{\% > 0} & \makecell[tl]{Mode \\(\% error)} & \makecell[tl]{$\SI{68}\percent$ quantile\\(\% error)} \\
\midrule
Linear   & URW & Ideal     & 52.8  & 0.07 & 1.99 \\
InceptNet & URW & Ideal     & 68.9  & 0.81 & 1.43 \\
ResNet    & URW & Ideal     & 71.2  & 0.70 & 1.52 \\
Linear   & URW & URW       & 32.4  & 0.89 & 8.92 \\
InceptNet & URW & URW       & 56.9  & 0.71 & 3.15\\
ResNet    & URW & URW       & 66.4  & 0.78 & 3.40 \\
\bottomrule
\end{tabular}
\caption{Quantitative properties of the fractional error histograms in Figure~\ref{fig:cluster_ideal_traindead} and 
\ref{fig:cluster_URW} }
\label{table:cluster_hist}
\end{centering}
\end{table}

\section{Systematic Variation Study}
\label{section:systematics}

We also tested the robustness of the CNN models and the linear algorithm to noise in the data. We applied noisy transformations to the input data, and tested the performance of the pre-trained models. We found that all models had predictable performance declines.

\subsection{Flat Noise}
\label{section:sys_flat}

The first test we tried was applying a flat 90\% or 110\% multiplier to all the charge in an image. This will necessarily shift the pre-trained linear algorithm's fractional error histogram by nearly the same amount. In theory, a CNN could adjust for this flat noise by considering the length of a shower in relation to its total charge. Additionally, in a dataset with background signals, the change in the background charge could serve as a reference point. However, our dataset had no background, and our CNNs' performance, plotted in Figure~\ref{fig:sys_noise} suffered the same bias shifts as the linear algorithm. 

\begin{figure}[h]
\subfloat{\includegraphics[width=.49\textwidth]{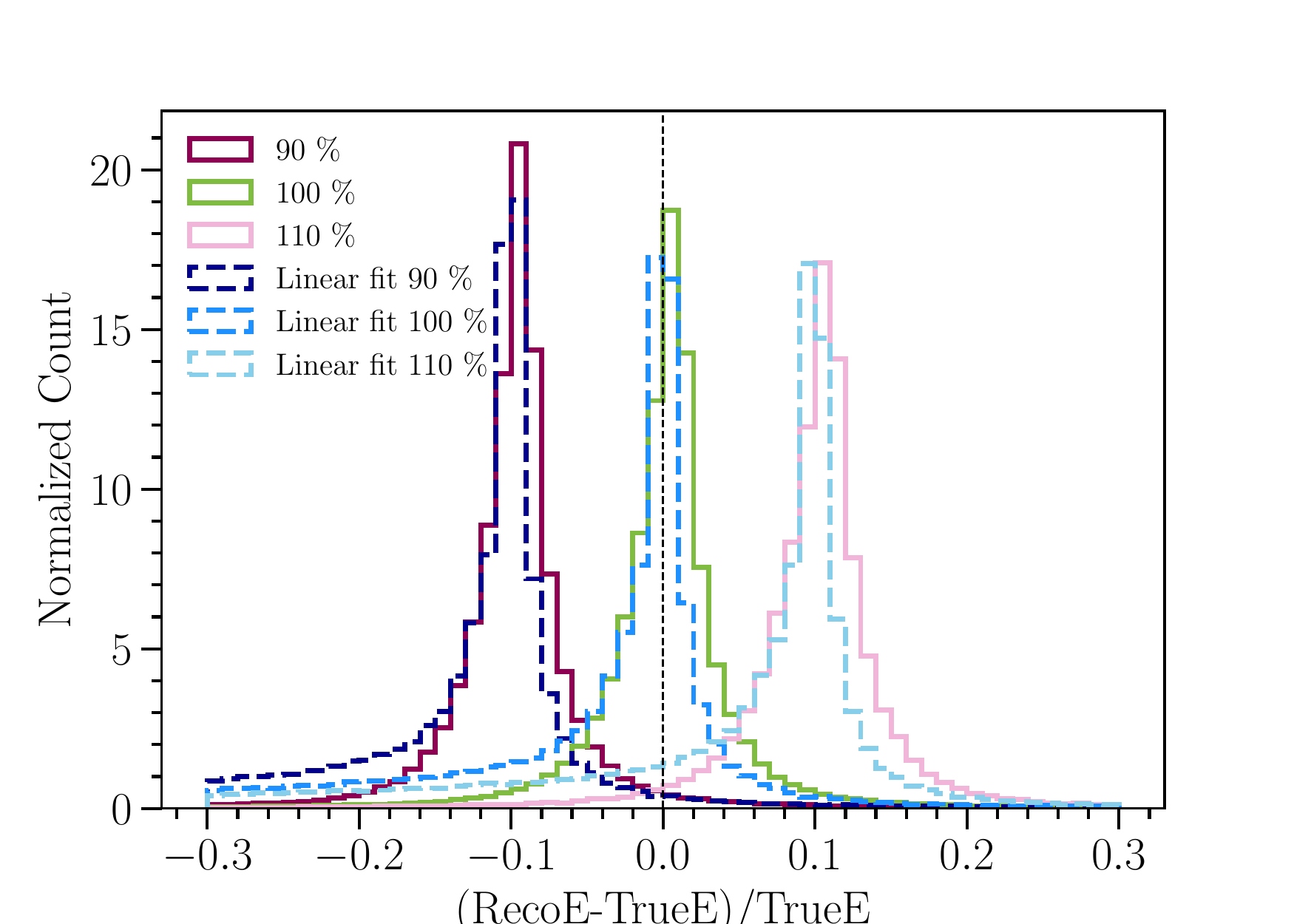}}
\subfloat{\includegraphics[width=.49\textwidth]{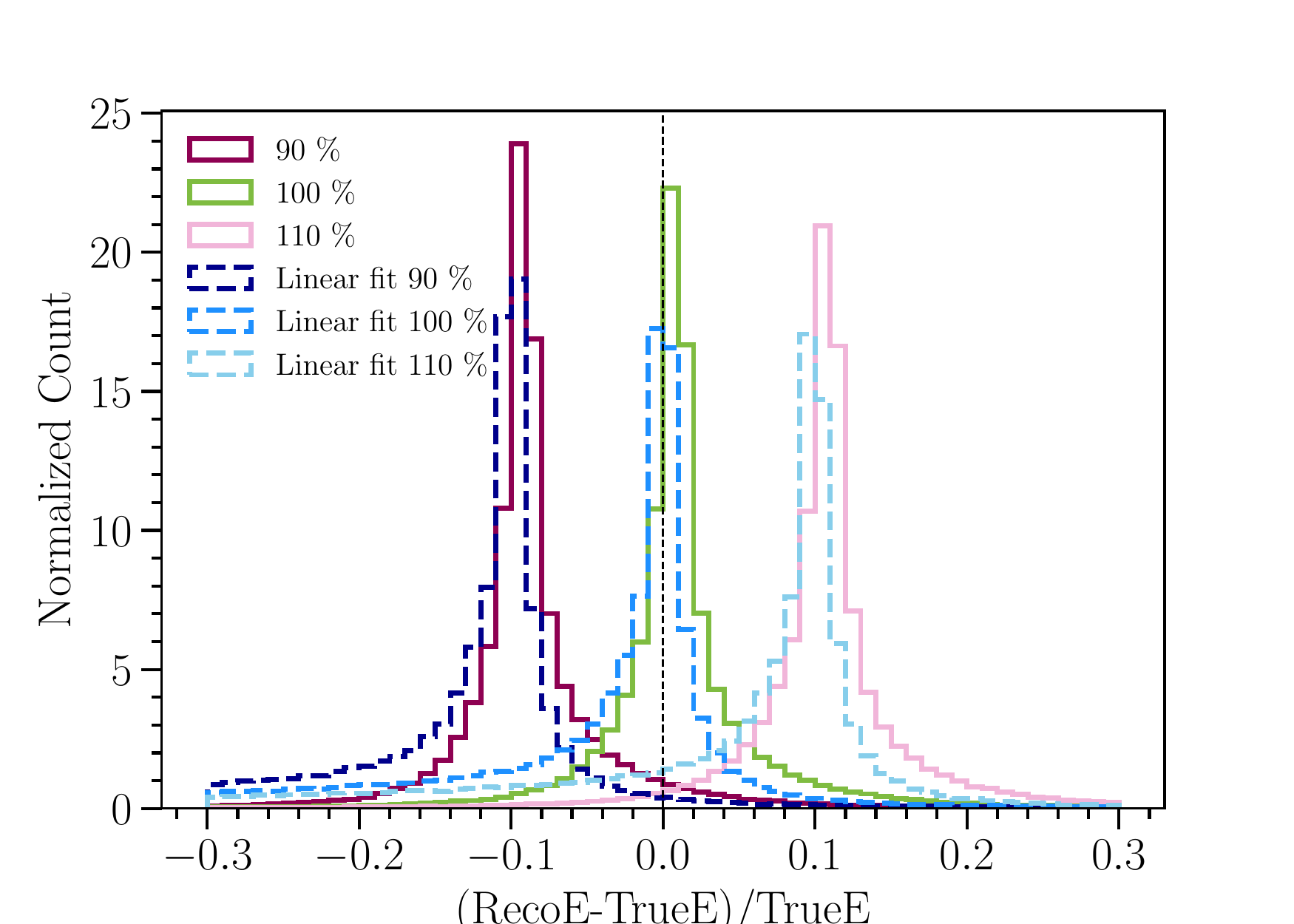}}
\caption{Histograms of the fractional error on input events with a 90\%, 100\%, and 110\% multiplier applied to the pixel charge Q, for the InceptNet (\textit{left}) and ResNet (\textit{right}) models, compared to the linear fit.}
\label{fig:sys_flat}
\end{figure}

\subsection{Gaussian Per-Wire Noise}
\label{section:sys_wire}
\begin{figure}[h]
\subfloat{\includegraphics[width=.49\textwidth]{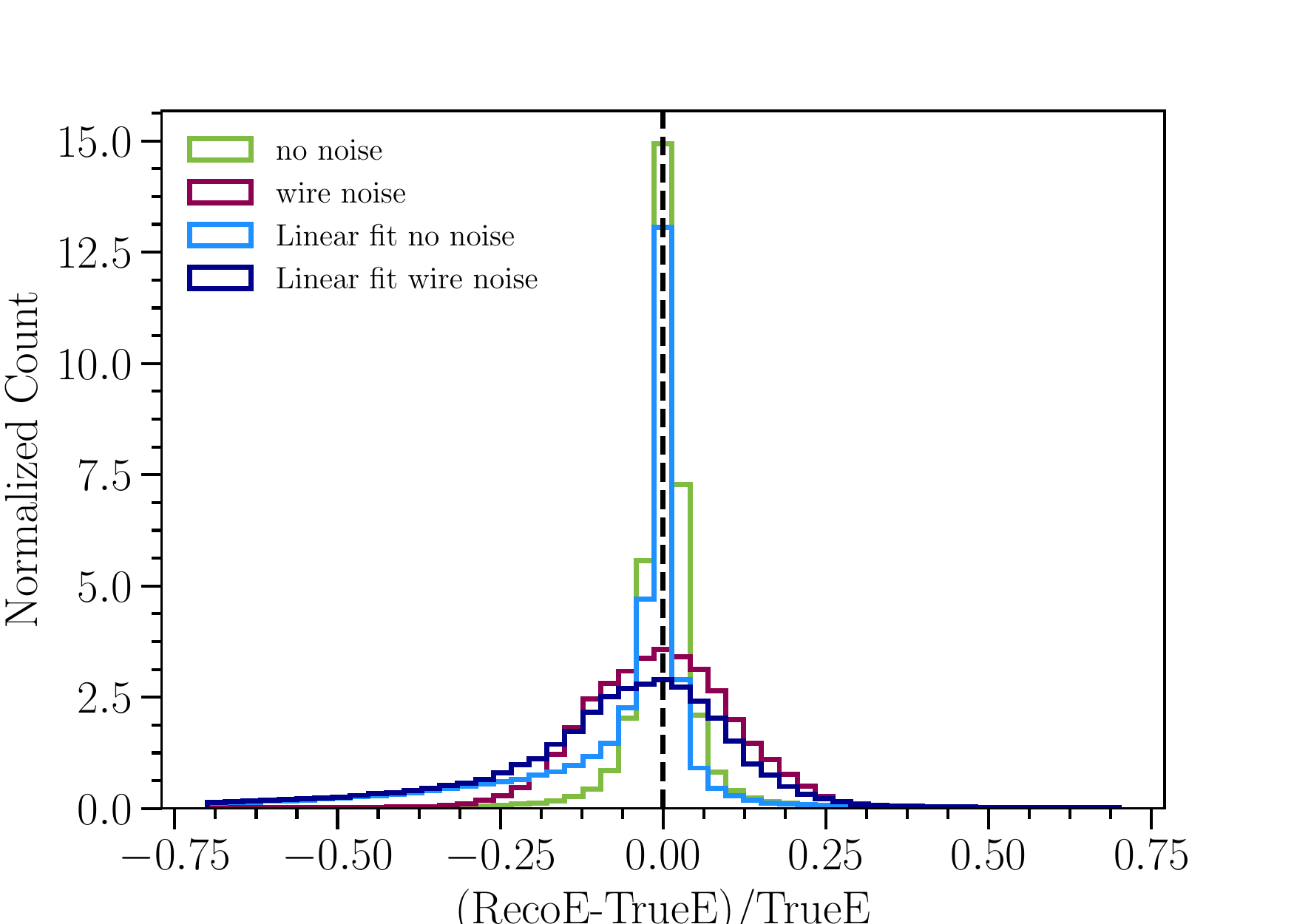}}
\subfloat{\includegraphics[width=.49\textwidth]{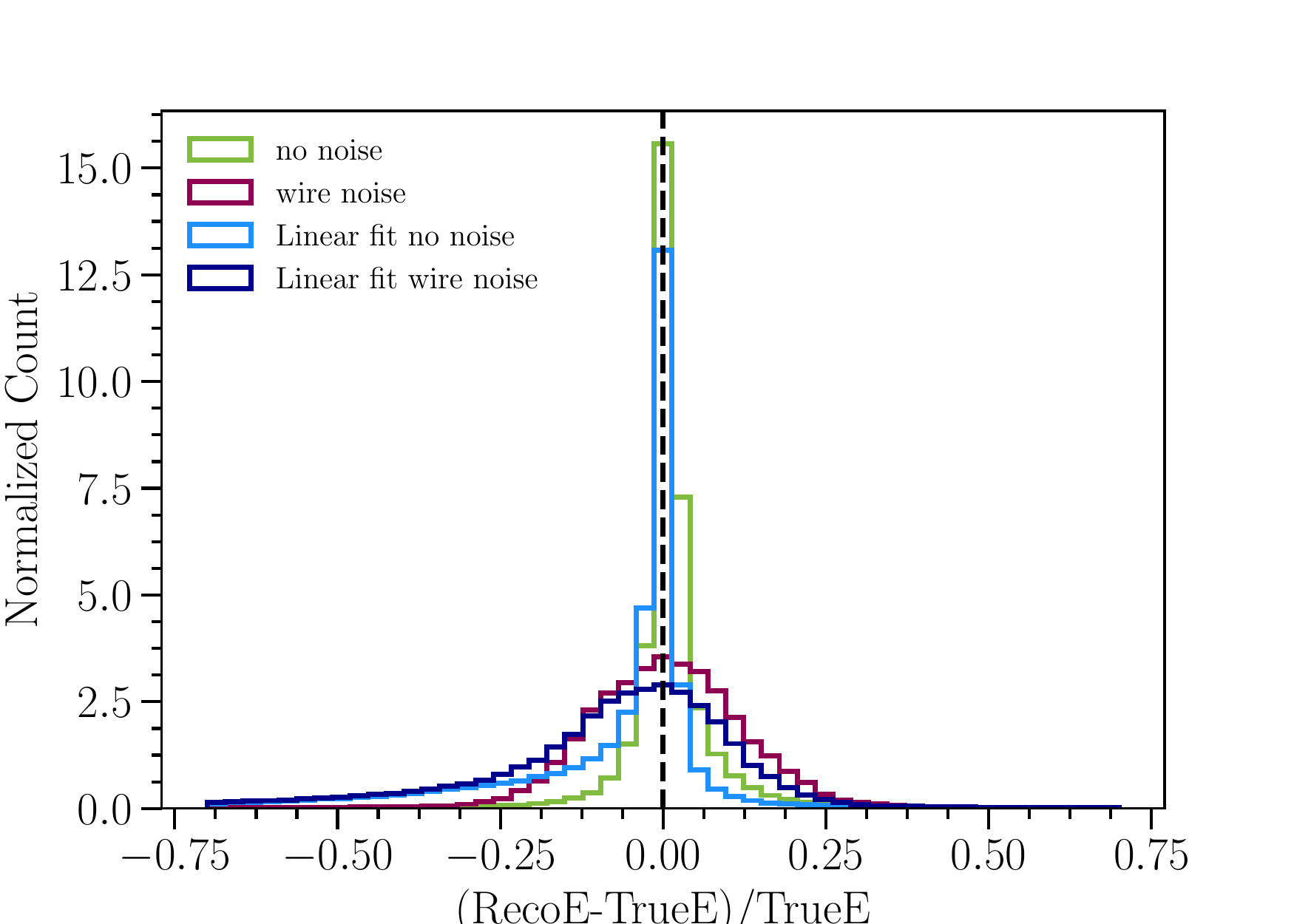}}
\caption{Histograms of the fractional error on input events with Gaussian distributed per-wire noise, for the InceptNet (\textit{left}) and ResNet (\textit{right}) models, compared to the linear fit.}
\label{fig:sys_noise}
\end{figure}

We also checked the models' robustness under Gaussian noise on the detector wires.
For this test, we multiplied each wire (column) in the input images by a random value pulled from a Gaussian distribution centered on 1 with width 0.1, and then evaluated the trained model's performance on this noisy dataset.
WE expected this kind of noise to increase the performance variance. This effect is what we see in Figure~\ref{fig:sys_noise}, where we plotted the resulting fractional error distribution. \\

On the dataset without noise, the linear algorithm predicts the top $\SI{68}\percent$ of shower energies to within $\SI{10.1}\percent$, and on that with noise it predicts them to within $\SI{16.5}\percent$.
The neural networks perform a little better: on the dataset without and with respectively, the InceptNet model predicts $\SI{68}\percent$ of events to within $\SI{3.2}\percent$ and $\SI{11.5}\percent$, while the ResNet model predicts them to within $\SI{3.4}\percent$ and $\SI{11.6}\percent$. 

\bibliographystyle{unsrt}
\bibliography{sorsamp}
\end{document}